\pdfoutput=1
\documentclass[aps,prb,twocolumn,superscriptaddress,amsmath]{revtex4-1}
\usepackage{graphicx}
\usepackage{amsmath}
\usepackage{subfigure}
\usepackage{epsfig}
\usepackage{float}

\begin{document}
\title{Electronic structures of [001]- and [111]-oriented InSb and GaSb free-standing nanowires}
\author {Gaohua Liao}
\affiliation{Key Laboratory for the Physics and Chemistry of Nanodevices and Department of Electronics, Peking University, Beijing 100871, China}
\affiliation{Department of Applied Physics and Key Laboratory for Micro-Nano Physics and Technology of Hunan Province, Hunan University, Changsha 410082, China}
\author {Ning Luo}
\affiliation{Key Laboratory for the Physics and Chemistry of Nanodevices and Department of Electronics, Peking University, Beijing 100871, China}
\author {Zhihu Yang}
\affiliation{Key Laboratory for the Physics and Chemistry of Nanodevices and Department of Electronics, Peking University, Beijing 100871, China}
\author {Keqiu Chen}
\affiliation{Department of Applied Physics and Key Laboratory for Micro-Nano Physics and Technology of Hunan Province, Hunan University, Changsha 410082, China}
\author {H. Q. Xu}
\email[Corresponding author. Electronic addresses: ]{hqxu@pku.edu.cn; hongqi.xu@ftf.lth.se}
\affiliation{Key Laboratory for the Physics and Chemistry of Nanodevices and Department of Electronics, Peking University, Beijing 100871, China}
\affiliation{Division of Solid State Physics, Lund University, Box 118, S-221 00 Lund, Sweden}

\date{\today}

\begin{abstract}
We report on a theoretical study of the electronic structures of InSb and GaSb nanowires oriented along the [001] and [111] crystallographic directions. The nanowires are described by atomistic, spin-orbit inteaction included, tight-binding models, and the band structures and the wave functions of the nanowires are calculated by means of a Lanczos iteration algorithm.  For the [001]-oriented InSb and GaSb nanowires, the systems with both square and rectangular cross sections are considered. Here, it is found that all the energy bands are double degenerate. Furthermore, although the lowest conduction bands in these nanowires show good parabolic dispersions, the top valence bands show rich and complex structures. In particular, the topmost valence bands of these nanowires with a square cross section show a double maximum structure. In the nanowires with a rectangular cross section, this double maximum structure is suppressed and top valence bands gradually develop into parabolic bands as the aspect ratio of the cross section is increased. For the [111]-oriented InSb and GaSb nanowires, the systems with hexagonal cross sections are considered. It is found that all the bands at the $\Gamma$-point are again double degenerate. However, some of them will split into non-degenerate bands when the wave vector moves away from the $\Gamma$-point, due to the presence of inversion asymmetry in the lattice structures. Furthermore, although the lowest conduction bands again show good parabolic dispersions, the topmost valence bands do not show the double maximum structure but, instead, a single maximum structure with its maximum at a wave vector slightly away from the $\Gamma$-point. The wave functions of the band states near the band gaps of the [001]- and [111]-oriented InSb and GaSb nanowires are also calculated and are presented in terms of probability distributions in the cross sections. It is found that although the probability distributions of the band states in the [001]-oriented nanowires with a rectangular cross section could be qualitatively described by a one-band effective mass theory, the probability distributions of the band states in the [001]-oriented nanowires with a square cross section and the [111]-oriented nanowires with a hexagonal cross section show characteristic patterns with symmetries being closely related to the irreducible representations of the relevant double point groups and do in general go beyond the prediction of a simple one-band effective mass theory. We also investigate the effects of quantum confinement on the band structures of the [001]- and [111]-oriented InSb and GaSb nanowires and present an empirical formula for the description of quantization energies of the band edge states in the nanowires, which could be used to estimate the enhancement of the band gaps of the nanowires as a result of quantum confinement.
\end{abstract}

\maketitle

\section{INTRODUCTION}

In recent years, InSb and GaSb nanowires have attracted great attention due to their unique electrical and optical properties and potential applications in nanoelectronics and optoelectronics.\cite{Chung-1,Kimberly-1,Caroff-1,Xu-9,Nilsson2010,Nilsson2011,Borg-1,Ganjipour2011,Thelander-1} These nanowires have also been considered as most suitable material systems for realization of topological superconducting nanowires\cite{Mourik2012,Xu-8,Churchill2013,Deng2014} in which Majorana bound states can be present at the ends of the nanowires\cite{Lutchyn2010,Oreg2010,Law2009,Flensberg2010,Sau2010,Stanescu2011,Huang2014} and could be exploited for quantum information processing.\cite{Navak2008} Bulk InSb and GaSb have narrow band gaps, small carrier effective masses and high carrier mobilities.  Especially, bulk InSb has a direct band gap of 0.17 eV, a small electron effective mass of 0.015 $m_e$ (where $m_e$ is the free electron mass), and a high electron mobility of 77000 cm$^2$/Vs, while bulk GaSb has a direct band gap of 0.73 eV and a high hole mobility of 1000 cm$^2$/Vs. High crystalline quality InSb and GaSb nanowires have been achieved through growth by molecular beam epitaxy (MBE),\cite{Thelander-1} chemical beam epitaxy (CBE)\cite{Ercolani-1,Vogel-1} and metal organic vapour phase epitaxy (MOVPE).\cite{Caroff-2,Jeppsson-1} Typical diameters of these epitaxially grown nanowires are in a range of a few nanometers to more than one hundred nanometers. These nanowires usually have zincblende lattice structures and can be grown along $<001>$, $<110>$ or $<111>$ crystallographic directions. It has also been demonstrated that it is possible to introduce well-controlled doping\cite{ZXYang-1,Borg-2} into the nanowires. High crystalline quality radial and axial nanowire heterostructures\cite{Borg-1,Ercolani-1,ZXYang-2} and superlattices\cite{Eltoukhy-1,Eltoukhy-2} have also been realized. In particular, nanowire heterostructures made with InSb and GaSb, in which the InSb segments show an n-type conduction behavior and the GaSb segments show a p-type conduction behavior, could be used to develop tunnel field-effect transistors, topological structures, and other novel devices and material systems. For a success in the development and optimization of novel devices and material systems made with InSb and GaSb nanowires as well as their heterostructures, it is inevitable to have a deep understanding of the electronic structures opf the InSb and GaSb nanowires.

Many theoretical studies of the electronic structures of semiconducting nanowires have been carried out using tight-binding methods.\cite{Xu-5,Xu-6,Xu-7,Xu-3,Xu-4,Niquet-1,Niquet-2} In comparison with other methods, such as density functional theory (DFT),\cite{Cahangirov-1,Keqiu-1} $\bf{k}\!\!\cdot\!\!\bf{p}$ theory,\cite{Lassen-1,Kishore-1,Kishore-2} and pseudopotential methods,\cite{WangLW-1}  tight-binding methods have been proved as more efficient and accurate methods for the calculation of the electronic structures of semiconductor nanowires with a diameter in a range of a few nanometers to more than 100 nanometers in the whole Brillouin zone.
Based on $sp^{3}s^{*}$ tight-binding methods,\cite{Vogl-1} Persson and Xu have studied the electronic structures of [100]-oriented InAs, InP, and GaAs nanowires with a square cross-section and rectangular cross section\cite{Xu-5,Xu-6,Xu-7} and of [111]-oriented GaAs and InP nanowires with a hexagonal cross section.\cite{Xu-3,Xu-4} Niquet \textit{et al.}\cite{Niquet-1} have studied the electronic structures of Si, Ge, GaP, InAs, InP and GaAs nanowires with a circular cross section using $sp^{3}$ or $sp^{3}d^{5}s^{*}$ tight-binding methods.\cite{Chadi-1,Jancu-1,Boykin-1} These studies have provided rich information for experimental studies of the electrical and optical properties of these semiconductor nanowires.

In this paper, we report on a theoretical study of the electronic structures of InSb and GaSb nanowires oriented along the [001] and [111] crystallographic directions using $sp^{3}s^{*}$ nearest-neighbor,  spin-orbit interaction included, tight-binding methods. We will show that all the energy bands of the [001]-oriented InSb and GaSb nanowires are double degenerate. Furthermore, although the lowest conduction bands show good parabolic dispersions, the top valence bands exhibit complex structures. In particular, the topmost valence bands of these nanowires with a square cross section show a double maximum structure. However, for the [001]-oriented InSb and GaSb nanowires with a rectangular cross section, the topmost valence bands develop into good parabolic bands as the aspect ration of the cross section increases. For the [111]-oriented InSb and GaSb nanowires, it is shown that although all the band states at the $\Gamma$ point are double degenerate, some of them will split into non-degenerate bands when the wave vector moves away from the $\Gamma$ point. Furthermore, the lowest conduction bands again show good parabolic dispersions, while the topmost valence bands show a single maximum structure with its maximum at a wave vector slightly shifted away from the $\Gamma$ point. Band edge energies have also been examined for the InSb and GaSb nanowires. In particular, we have fitted the band edge energies as a function of the lateral size of these nanowires to a simple formula with witch the quantization energies in the nanowires can be quickly estimated. In addition, the wave functions of the band states near the band gaps of the InSb and GaSb nanowires are calculated. We will show that although the probability distributions of the band states in the [001]-oriented nanowires with a rectangular cross section could be qualitatively described by a one-band effective mass theory, the probability distributions of the band states in the [001]-oriented nanowires with a square cross section and the [111]-oriented nanowires with a hexagonal cross section show characteristic patterns with symmetries being closely related to the irreducible representations of the relevant double point groups and could in general not be reproduced by the calculations based on a simple one-band effective mass theory. 

The rest paper is organized as follows. In Section II, the formalism is presented together with a brief description of a Lanczos iteration algorithm employed for the calculations of the band structure and the wave functions of electronic states located near the band gap of a semiconductor nanowire. In Section III, the calculated electronic structures of [001]-oriented InSb and GaSb nanowores with both a square cross section and a rectangular cross section are presented and discussed. Section IV is devoted to the description and discussion of the calculated electronic structures of [111]-oriented InSb and GaSb nanowires with a hexagonal cross section. Finally, the paper is summarized in Section V.

\section{Method of Calculations}

\subsection{Tight-binding formalism}

In this paper, the $sp^{3}s^{*}$ nearest-neighbor, spin-orbit interaction included, tight-binding formalism is employed in the calculation of the electronic structures of [001]- and [111]-oriented free-standing InSb and GaSb nanowires. In a tight-binding formalism, Bloch sums of the form\cite{Carlo-1}
\begin{equation}
\label{eq01}
|\alpha,\nu,\bf{k}\rangle=\frac{1}{\sqrt{N}}\sum_{R} e^{i\bf{k}\cdot\bf{R}_{\nu}}|\alpha,\bf{R}_{\nu}\rangle ,
\end{equation}
are used as a basis, where $\bf{R}_{\nu}=\bf{R}+\bf{r}_{\nu}$ with $\bf{R}$ standing for a lattice vector and $\bf{r}_{\nu}$ an atomic displacement vector in a unit cell, $\bf N$ is the number of lattice sites, and $|\alpha,\bf{R}_{\nu}\rangle$ stands for an atomic orbital $\alpha$ at $\bf{R}_{\nu}$. In the $sp^{3}s^{*}$ nearest-neighbor, spin-orbit interaction included, tight-binding formalism, the atomic orbitals are chosen as 10 localized, spin-dependent orbitals $|s\uparrow\rangle$, $|p_{x}\uparrow\rangle$, $|p_{y}\uparrow\rangle$, $|p_{z}\uparrow\rangle$, $|s^{*}\uparrow\rangle$, $|s\downarrow\rangle$, $|p_{x}\downarrow\rangle$, $|p_{y}\downarrow\rangle$, $|p_{z}\downarrow\rangle$, and $|s^{*}\downarrow\rangle$. In the basis of the Bloch sums, the Hamiltonian $\bf{H}$ can be written in a matrix form with matrix elements given by
\begin{equation}
\label{eq02}
\bf{H}_{\alpha\nu,\beta\xi}(k)=\sum_{R}e^{-i\bf{k}\cdot(\bf{R}_{\nu}^{\prime}-\bf{R}_{\xi})}\langle{\alpha,\bf{R}_{\nu}^{\prime}|} \bf{H}|{\beta,\bf{R}_{\xi}}\rangle ,
\end{equation}
and the eigenfunctions in a form of
\begin{equation}
\label{eq03}
{|n,\bf{k}}\rangle=\sum_{\alpha,\nu}c_{n,\alpha\nu}|{\alpha,\nu,\bf{k}}\rangle .
\end{equation}

\begin{figure}[t]
\begin{center}
\includegraphics[width=85mm]{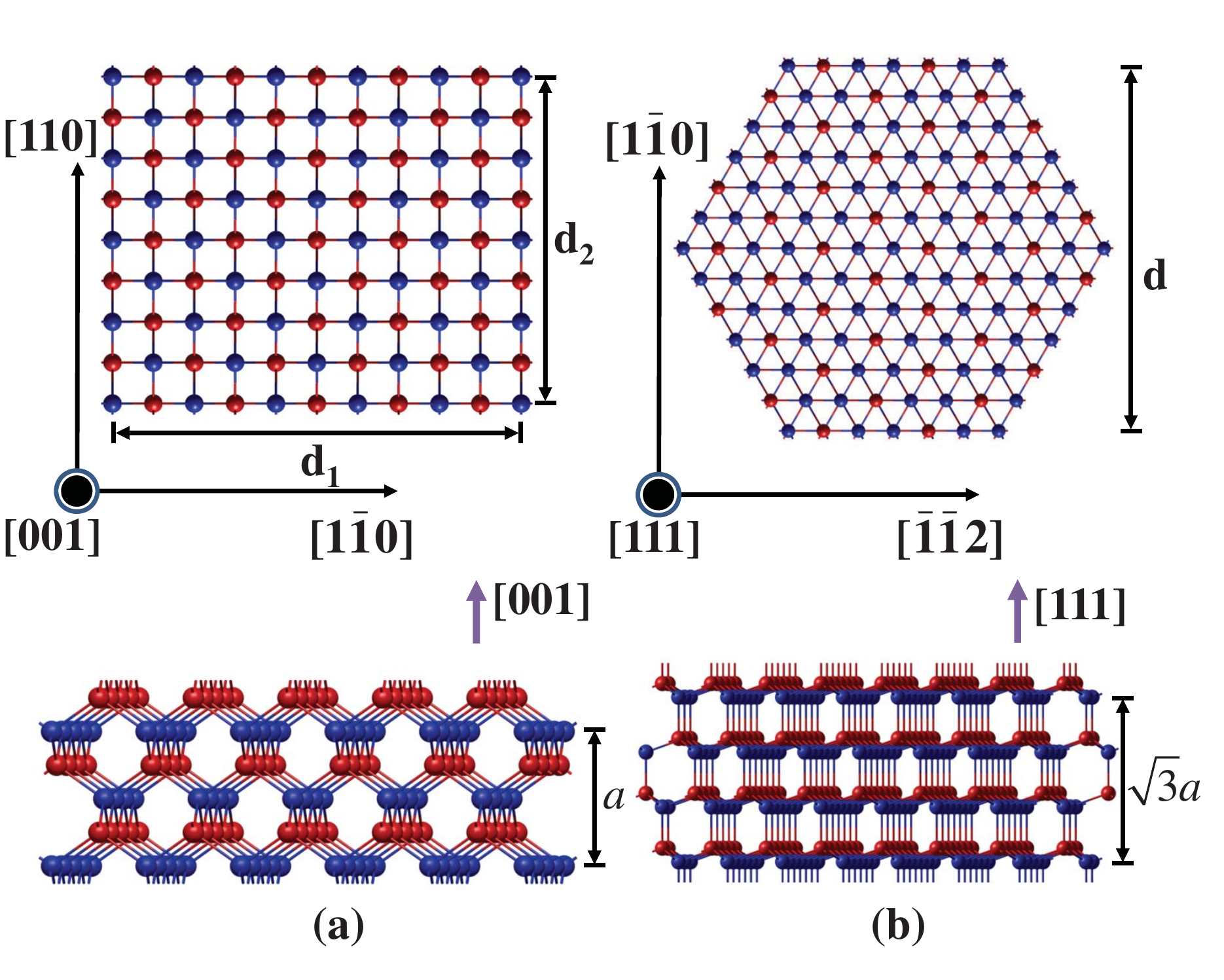}
\caption{(Color online) Atomistic tight-binding model structures employed in this work. (a) Top view and side view of a unit cell of a [001]-oriented nanowire with lateral size of $d_{1}\times d_{2}$. (b) Top view and side view of a unit cell of a [111]-oriented nanowire with lateral size d. Blue and red spheres represent cation (In or Ga) atoms and anion (Sb) atoms, respectively. The surface dangling bonds of the nanowires are passivated using H atoms which are not shown in these figures.}
\label{Fig:model}
\end{center}
\end{figure}

\subsection{Nanowire models}

Free-standing nanowires considered in this work are zincblende crystals, oriented along the [001] and [111] crystallographic directions with $\{110\}$ facets. Figure~\ref{Fig:model} shows atomistic model structures of unit cells of a [001]- and a [111]-oriented zincblende crystalline nanowire.  As shown in the figure, a [001]-oriented nanowire is assumed to have a square or a rectangular cross section, while a [111]-oriented nanowire is to have a hexagonal cross section. The periods of the [001]- and [111]-oriented  nanowires are $a$ and $\sqrt{3}a$, respectively, where $a$ is the lattice constant of the corresponding bulk material.  Thus, the Brillouin zone is defined as $-2\pi/a \le k \le 2\pi/a$ for a nanowire oriented along the [001] direction and as $-2\pi/(\sqrt{3}a)\le k\le 2\pi/(\sqrt{3}a)$ for a nanowire oriented along the [111] direction. In the calculations for the electronic structures, surface dangling bonds of the nanowires have been passivated using hydrogen atoms. The on-site energies of the hydrogen atoms are determined from the $s$-orbital energies of the neutral hydrogen atoms and the $s$-orbital energies of the nanowire host atoms using a scaling method of Vogl \textit{et al.}\cite{Vogl-1} The interaction parameters between the hydrogen atoms and the passivated nanowire host surface atoms are determined using the Wolfsberg-Helmholz formula\cite{Wolf-1} in a procedure described by Xu and Lindefelt\cite{Xu-1,Xu-2} by setting the distances between the hydrogen atoms and their passivated nanowire surface atoms to an optimal value of $0.4l$, where $l$ is the bond length of the host bulk material.  On-site energies of hydrogen atoms and the interaction parameters between the hydrogen atoms and their passivated nanowire surface atoms determined for this work are given in Table~\ref{tab:Table1}, where the tight-binding parameters of the bulk materials of Ref.~\onlinecite{Klimech-1} employed in this work are also included for completeness and convenience.  With these hydrogen-related atomic tight-binding parameters, the surface states of the nanowires are pushed to locate in energy far remote from the fundamental band gaps. Thus, the effects of the dangling bonds on the band states near the fundamental 
band gaps are eliminated in this work.

\begin{table}[t]
\caption{Tight-binding parameters employed in this work in units of eV. The on-site energies of the hydrogen atoms are determined using a scaling method of Ref.~\onlinecite{Vogl-1}. The interaction parameters between the passivation H atoms and the passivated nanowire host surface atoms are derived using a procedure described in Ref.~\onlinecite{Xu-1}, in which the distance between each H atom and its passivated host atom is taken as $0.4l$, where $l$ is the bond length of the relevant bulk material. The material parameters for bulk InSb and GaSb taken from Ref.~\onlinecite{Klimech-1} are also included in the table for completeness.}
\begin{center}
\begin{tabular*}{80mm}{@{\extracolsep{\fill}}c c c}
\hline\hline
Parameter   &InSb     &GaSb \\
\hline
$E_{sa}$                     &-7.80905   &-7.16208\\
$E_{pa}$                     &-0.14734   &-0.17071\\
$E_{sc}$                      &-2.83599   &-4.77036\\
$E_{pc}$                     &3.91522    &4.06643\\
$E_{s^{*}a}$              &7.43195    &7.32190\\
$E_{s^{*}c}$            &3.54540    &3.12330\\
$V_{ss}$                  &-4.89637   &-6.60955\\
$V_{xx}$                 &0.75260    &0.58073\\
$V_{xy}$                &4.48030    &4.76520\\
$V_{sapc}$           &3.33714    &3.00325\\
$V_{scpa}$           &5.60426    &7.78033\\
$V_{s^{*}apc}$    &4.59953    &4.69778\\
$V_{pas^{*}c}$   &-2.53756   &4.09285\\
$\Delta_{a}$        &0.85794    &0.75773\\
$\Delta_{c}$        &0.51000       &0.15778\\
$E_{sH}$            &-5.86343   &-5.91734 \\
$V_{sasH}$        &-8.26511        &-10.1974\\
$V_{scsH}$       &-7.39861        &-9.36432\\
$V_{sHpa}$     &8.80883     &10.4230\\
$V_{sHpc}$     &3.52086  &2.84691\\

    \hline \hline
\end{tabular*}
\end{center}
 \label{tab:Table1}
\end{table}

\begin{figure*}[t]
    \begin{center}
\includegraphics[width=7in]{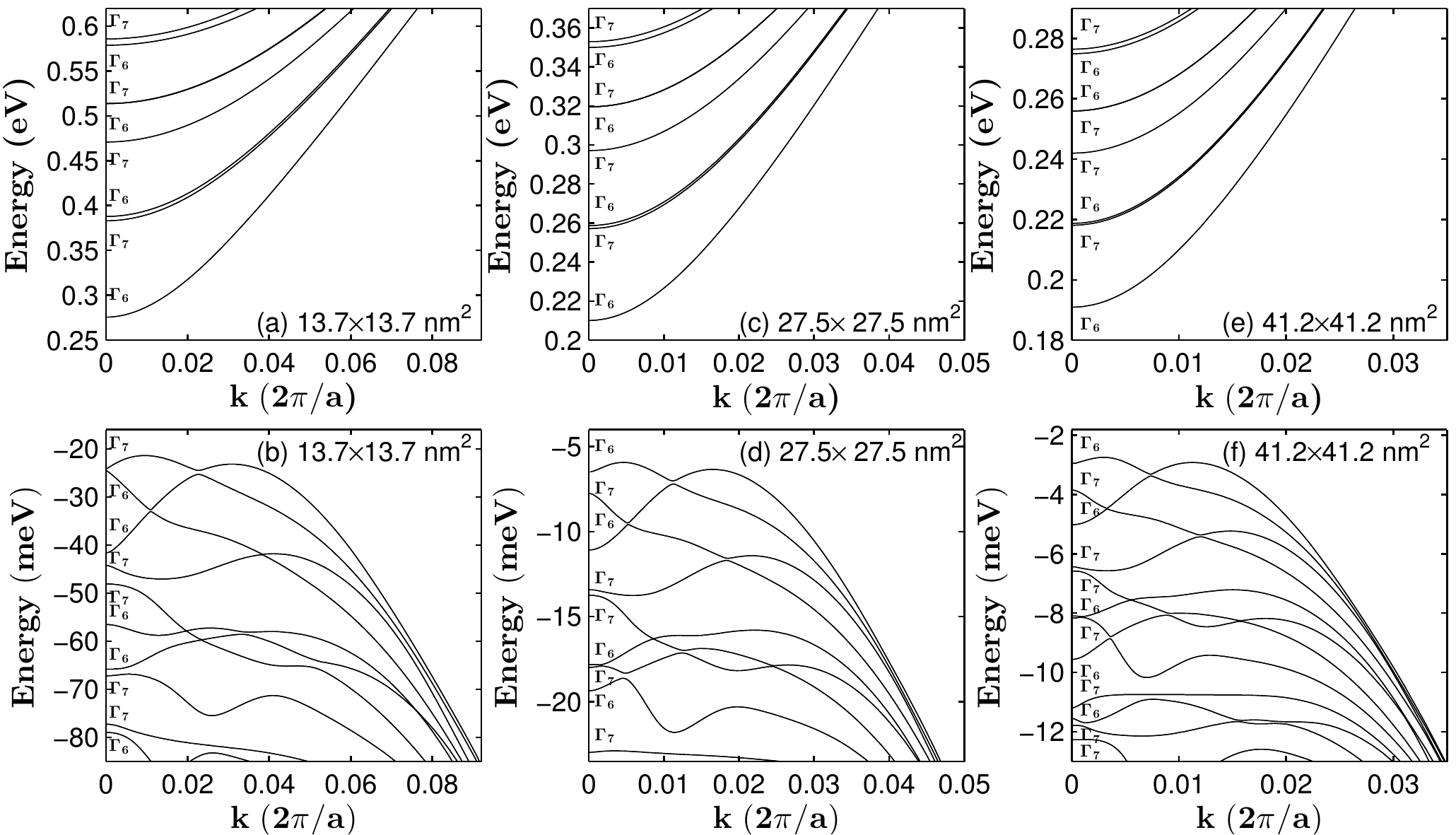}
       \caption{Band structures of [001]-oriented InSb nanowires with a square cross section of the lateral sizes of (a) and (b) 13.7$\times$13.7 nm$^{2}$, (c) and (d) 27.5$\times$27.5 nm$^{2}$, and (e) and (f) 41.2$\times$41.2 nm$^{2}$. The bands are labeled by the symmetries of their band states at the $\Gamma$-point in correspondence with the irreducible representations, $\Gamma_{6}$ and $\Gamma_{7}$, of the $D_{2d}$ double point group. All the bands are double degenerate.}
       \label{Fig:[001]InSbbandsq}
       \end{center}
\end{figure*}

\subsection{The Lanczos iteration method}

In a semiconductor nanowire, unit cells normally contain an extremely large number of atoms. For example, in a [111]-oriented nanowire with a lateral size of 30 nm, the number of atoms in a unit cell is about 35000 atoms.  In the $sp^{3}s^{*}$ tight-binding formalism, the resulting Hamiltonian matrix is then on the order of $350000\times 3500000$, which is too large to be solved by direct diagonalization methods. To get the eigensolutions of such a large sparse matrix, we use the Lanczos iteration method. The Lanczos iteration method is a powerful method for solving the eigenvalue problem of a sparse Hamiltonian matrix and the method transforms the given Hamiltonian matrix to a tridiagonal Hamiltonian matrix with a small computational effort. However, because of numerical rounding errors, the Lanczos vectors generated during the iteration will lose their orthogonality. This problem could be handled by employing the following two numerical schemes with additional but reasonably small computational efforts: performing a full reorthogonalization of the generated Lanczos vectors after each step of the Lanczos iteration and computing desired eigenvectors by an inverse iteration procedure.\cite{Golub-1,Xu-6} In this paper, the former scheme has been used in most calculations, while the latter scheme is used only when the eigensolutions of the nanowires with a very large lateral size are computed.

In order to achieve fast convergence, the symmetry properties of the nanowires have been exploited in the calculations of the eigensolutions of the nanowires. This has been implemented by constructing symmetric starting Lanczos vectors using the projection operators of the relevant point groups, \cite{Melvin-1,Tinkham-1}
\begin{equation}
\label{eq04}
P_{kk}^{(j)}=\frac{l_{j}}{h}\sum_{R}\Gamma^{(j)}(R)^{*}_{kk}R ,
\end{equation}
where $\Gamma^{j}(R)_{kk}$ is the $k$th diagonal element in the $j$th irreducible representation of the symmetric operation $R$, $l_j$ is the dimentionality of the $j$th representation and $h$ is the number of the symmetric operations in a point group. The projection operator projects out the part of a given vector $|\phi_n>$ belonging to the $k$th row of the $j$th representation,
\begin{equation}
|\phi^{(j)}_{n,k}\rangle =P_{kk}^{(j)}|\phi_n\rangle .
\end{equation}
The symmetric Lanczos starting vectors can then be chosen as
\begin{equation}
\label{eq05}
|\Phi_{k}^{(j)}\rangle =B \sum_{n}C_n |\phi^{(j)}_{n,k}\rangle ,
\end{equation}
where $B$ is the normalization constant and $C_n$ is used to shape the starting vectors and has been set to unity in this work.

\begin{figure*}[t]
\centering
   \includegraphics[width=7in]{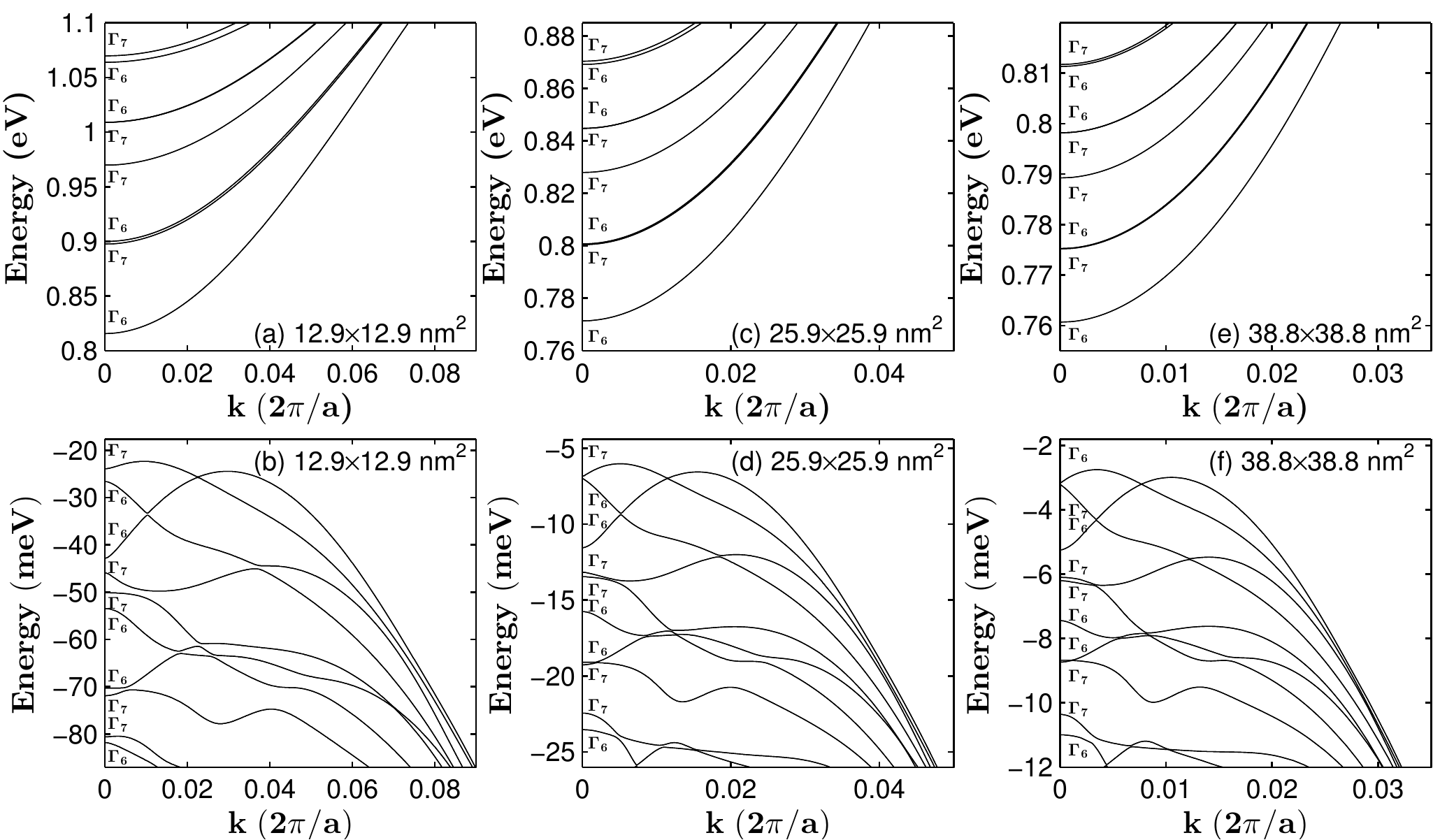}
\caption{Band structures of [001]-oriented GaSb nanowires with a square cross section of the lateral size of (a) and (b) 12.9$\times$12.9 nm$^{2}$, (c) and (d) 25.9$\times$25.9 nm$^{2}$,  and (e) and (f) 38.8$\times$38.8 nm$^{2}$. The bands are labeled by the symmetries of their band states at the $\Gamma$-point in correspondence with the irreducible representations, $\Gamma_{6}$ and $\Gamma_{7}$, of the $D_{2d}$ double point group. All the bands are double degenerate.}
\label{Fig:[001]GaSbsq}
\end{figure*}

\section{Electronic Structures of [001]-oriented I\lowercase{n}S\lowercase{b} and G\lowercase{a}S\lowercase{b} nanowires}

In this section, we report on the electronic structures of [001]-oriented InSb and GaSb nanowires with square and rectangular cross sections of different sizes. Band structures and wave functions of the nanowires are calculated and their symmetric properties are investigated. In a [001]-oriented nanowire with a square (rectangular) cross section, the crystallographic structure is symmetric under the operations of the $D_{2d}$ ($C_{2v}$) point group. At the $\Gamma$ point ($k=0$), the band structure Hamiltonian of the nanowire has the same symmetry as the nanowire crystallographic structure. However, due to the inclusion of spin-orbit interaction in the Hamiltonian, the symmetric properties of the electronic structure need to be characterized by the corresponding double point group. The $D_{2d}$ double point group has two double degenerated, double-valued irreducible representations $\Gamma_{6}$ and $\Gamma_{7}$, while the $C_{2v}$ double point group has only one doubly degenerate, double-valued irreducible representation $\Gamma_{5}$. Thus, for a nanowire with a square cross section, the energy bands at the $\Gamma$ point are all doubly degenerated and can be labeled by $\Gamma_{6}$ and $\Gamma_{7}$ according to their symmetries. For a nanowire with a rectangular cross section, all the energy bands at the $\Gamma$ point are again double degenerate, but are $\Gamma_5$-symmetric. At a finite value of $k$, both the band structure Hamiltonian of a nanowire with a square cross section and the band structure Hamiltonian of a nanowire with a rectangular cross section are symmetric under the operations of the $C_{2v}$ point group and therefore all the energy band states are double degenerate and are $\Gamma_5$-symmetric.

\subsection{Band structures of [001]-oriented InSb and GaSb nanowires}

Figure~\ref{Fig:[001]InSbbandsq} shows the calculated band structures of the [001]-oriented InSb nanowires with a square cross section of sizes 13.7$\times$13.7 nm$^{2}$, 27.5$\times$27.5 nm$^{2}$, and 41.2$\times$41.2 nm$^{2}$, while Fig.~\ref{Fig:[001]GaSbsq} shows the calculated band structures of the [001]-oriented GaSb nanowires with a square cross section of sizes 12.9$\times$12.9 nm$^{2}$, 25.9$\times$25.9 nm$^{2}$, and 38.8$\times$38.8 nm$^{2}$. The bands are labeled according to their state symmetries at $\Gamma$-point, namely, the double-valued irreducible representations of the $D_{2d}$ point group. It is seen that for both the InSb and the GaSb nanowires, the band gaps are increased as the lateral sizes decrease as a result of quantum confinement. Furthermore, the lowest conduction bands of the InSb and GaSb nanowires show simple, good parabolic dispersions around the $\Gamma$ point. These bands also have the same symmetry ordering except for the 5th and 6th lowest conduction bands which are nearly degenerate and may show different symmetry orderings at different sizes. In addition, the energy separation between the second and the third lowest conduction band is small and the two bands become non-distinguishable when the nanowire size becomes sufficiently large. However, the valence bands of these nanowires are complicated and show complex characteristics, such as non-parabolic dispersions and anti-crossings between the bands. In particular,  the topmost valence bands of these nanowires show a double-maximum structure--a characteristic which has also been found in the calculations for the band structures of InAs, InP, and GaAs nanowires.\cite{Xu-5,Xu-7} Also, the symmetry ordering of the two highest valence band states at the $\Gamma$-point is size-dependent. For example, in the InSb nanowire with the lateral size of 13.7$\times$13.7 nm$^{2}$, the highest valence band state at the $\Gamma$-point is $\Gamma_7$-symmetric and the second highest valence band state at the $\Gamma$-point is $\Gamma_6$-symmetric. When the lateral size of the InSb nanowires is increased to 27.5$\times$27.5 nm$^{2}$, the symmetry ordering of these two highest valence band states is reversed. The same symmetry ordering reversing with increasing lateral size has also been found in the two highest valence bands of the [001]-oriented GaSb nanowires with a square cross section. However, as shown in Fig.~\ref{Fig:[001]GaSbsq}, this symmetry ordering reversing occurs in a larger size.

\begin{figure}[t]
 \centering
 \includegraphics[width=85mm]{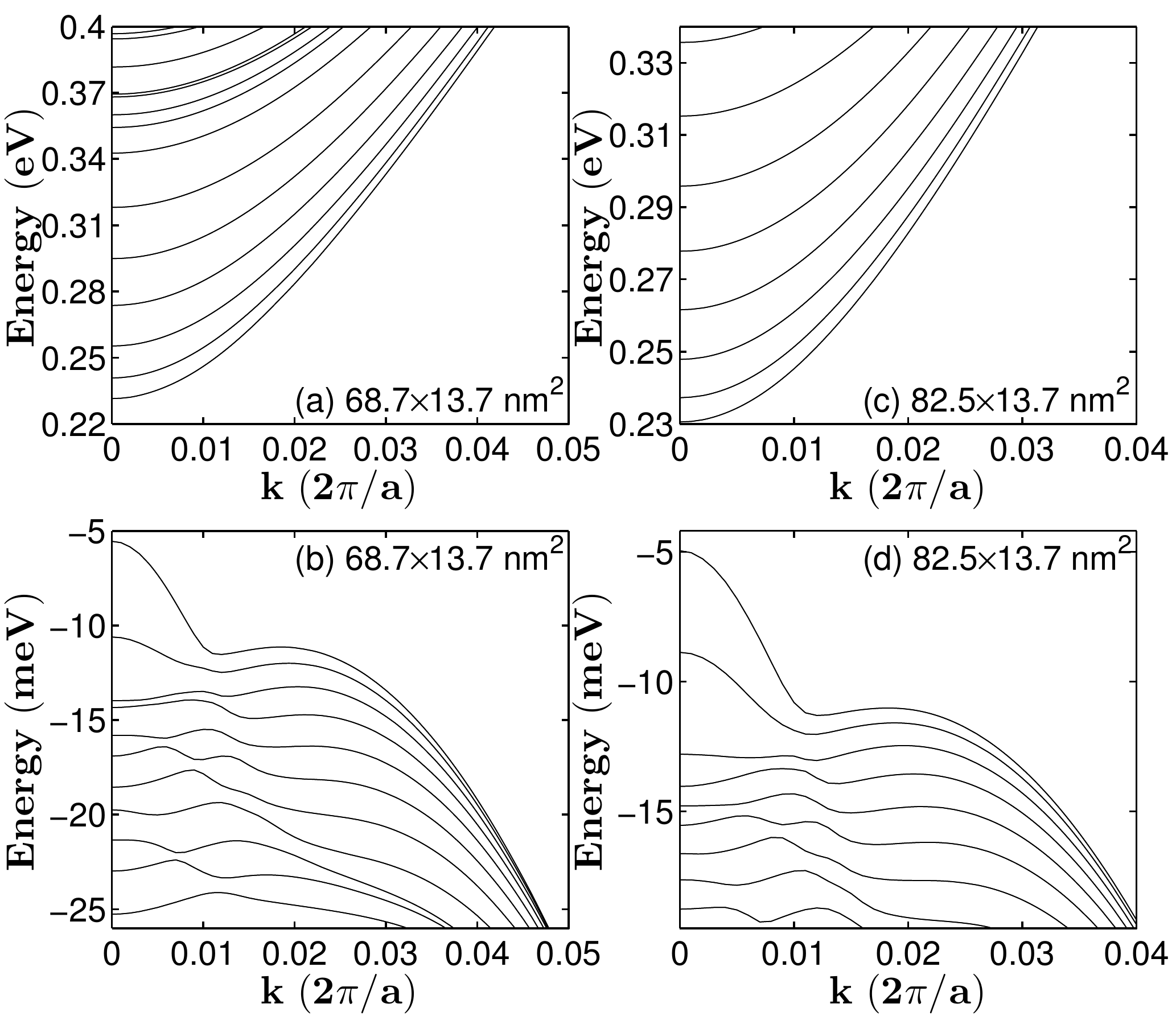}
       \caption{Band structures of [001]-oriented InSb nanowires with a rectangular cross section of the lateral sizes of (a) and (b) $68.7\times13.7$ nm$^{2}$, and (c) and (d) $82.5\times13.7$ nm$^{2}$. All the bands are $\Gamma_5$-symmetric and are double degenerate.}
       \label{Fig:[001]InSbbandrec}
\end{figure}

\begin{figure}[t]
 \centering
   \includegraphics[width=85mm]{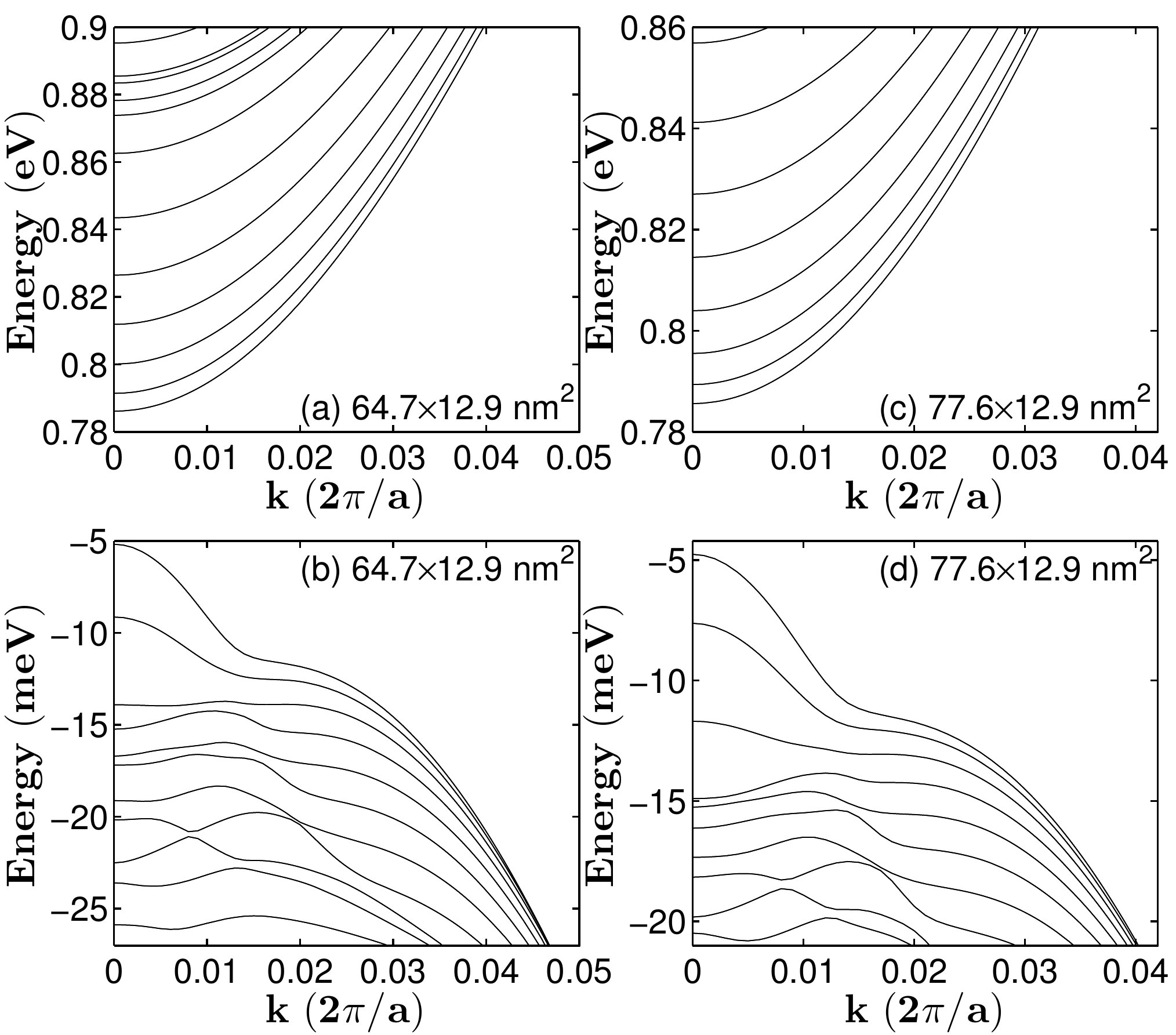}
       \caption{Band structures of [001]-oriented GaSb nanowires with a rectangular cross section of the lateral sizes of (a) and (b) $64.7\times12.9$ nm$^{2}$, and (c) and (d) $77.6\times12.9$ nm$^{2}$. All the bands are $\Gamma_5$-symmetric and are double degenerate.}
        \label{Fig:[001]GaSbbandrec}
\end{figure}

Now we study how the band structure develops when the cross section of a [001]-oriented nanowire becomes rectangular and the aspect ratio $d_1/d_2$ of the two rectangular sides $d_1$ and $d_2$ is changed. Figure~\ref{Fig:[001]InSbbandrec} shows the band structures of the [001]-oriented InSb nanowires with a rectangular cross section of sizes 68.7$\times$13.7 nm$^{2}$ and 82.5$\times$13.7 nm$^{2}$ and Fig.~\ref{Fig:[001]GaSbbandrec} shows the band structures of the [001]-oriented GaSb nanowires with a rectangular cross section of sizes 64.7$\times$12.9 nm$^{2}$ and 77.6$\times$12.9 nm$^{2}$. As we already mentioned, the band structure Hamiltonians of these nanowires are symmetric under the operations of the $C_{2v}$ point group. Thus, all the bands are double degenerate and are $\Gamma_{5}$-symmetric, and no band crossings could occur in the band structures. In particular, the lowest conduction bands of the rectangular nanowires show good parabolic dispersions around the $\Gamma$-point and, in diffrence from the corresponding nanowires with a square cross section, these bands are all well separated in energy. The valence bands of the rectangular nanowires also tend to show much simpler dispersions. In particular, the double-maximum structure found in the topmost valence band of an InSb nanowire or a GaSb nanowire has been suppressed with increasing aspect ratio $d_1/d_2$ of the rectangular cross section. In fact, the top valence bands of an InSb or a GaSb nanowire with a rectangular cross section become more parabolic around the $\Gamma$-point and will eventually recover the parabolic band properties of the corresponding quantum well system with increasing aspect ratio of the rectangular cross section. Such a development in the valence band structure has also been seen in our early study for [001]-oriented InAs, InP and GaAs nanowires.\cite{Xu-5,Xu-7}

\begin{table*}[t]
\caption{Parameters $p_{1}$, $p_{2}$, $p_{3}$, and $p_{4}$ in Eq.~(\ref{eq06}) obtained by fitting the equation to the calculated energies at the $\Gamma$-point of the lowest conduction band  and the highest valence band of the [001]-oriented InSb and GaSb nanowires with a squear cross section (labeled by subscript ``{\em squ}") and with a rectangular cross section (labeled by subscript ``{\em rec}").}
\begin{center}
\begin{tabular*}{140mm}{@{\extracolsep{\fill}}c c c c c c c }
\hline\hline
Material & Nanowire & Band shift & $p_{1}$             &$p_{2}$             &$p_{3}$     &$p_{4}$\\
         &  type    &    (eV)    &(eV$^{-1}$nm$^{-2}$) &(eV$^{-1}$nm$^{-1}$)&(eV$^{-1}$) &(eV)   \\
\hline
\qquad   &\!\!$[001]_{squ}$  &\!$\Delta E_{c}$  &\!0.01540  &0.45613  &0.22507 &\qquad\\
InSb     &\!\!$[001]_{squ}$  &\!$\Delta E_{v}$  &\!-0.15612  &-1.00194  &0.38529 &\qquad\\
\qquad   &\!\!$[001]_{rec}$  &\!$\Delta E_{c}$  &\!0.05352  &0.71705  &1.08493 &0.05914\\
\qquad   &\!\!$[001]_{rec}$  &\!$\Delta E_{v}$  &\!-0.08067  &-1.00953  &-19.13721 &-0.00342\\
\hline
\qquad   &\!\!$[001]_{squ}$  &\!$\Delta E_{c}$  &\!0.05797  &0.36375  &1.08645 &\qquad\\
GaSb     &\!\!$[001]_{squ}$  &\!$\Delta E_{v}$  &\!-0.18980  &-0.80449  &0.23092 &\qquad\\
\qquad   &\!\!$[001]_{rec}$  &\!$\Delta E_{c}$  &\!0.12114  &0.77146  &1.59858 &0.03330\\
\qquad   &\!\!$[001]_{rec}$  &\!$\Delta E_{v}$  &\!-0.17555  &0.45421  &-26.31477 &-0.00382\\
    \hline \hline
\end{tabular*}
\end{center}
\label{tab:Table2}
\end{table*}

\begin{figure}[t]
 \centering
\includegraphics[width=85mm]{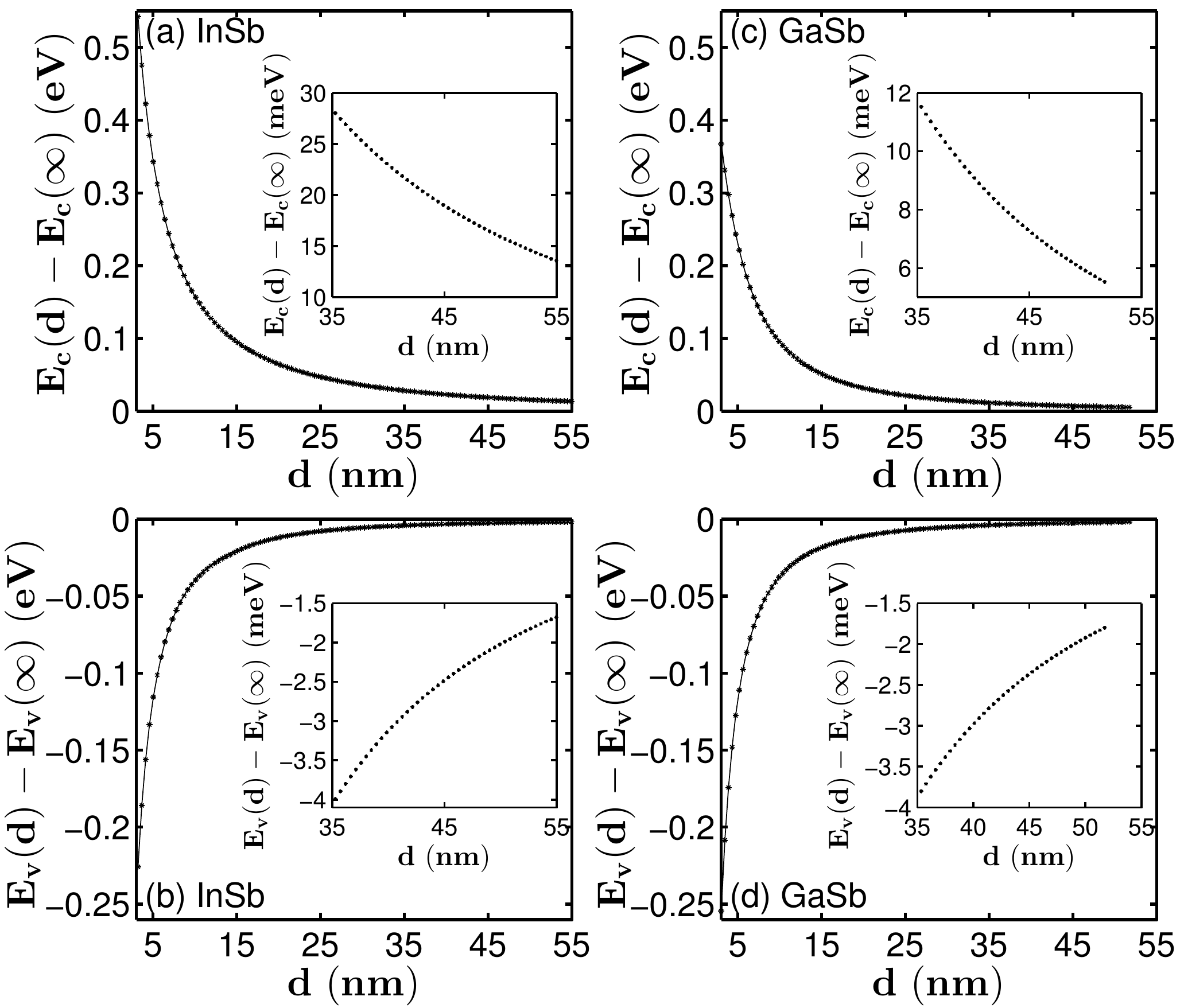}
      \caption{Lowest conduction band electron and highest valence band hole confinement energies in the [001]-oriented InSb and GaSb nanowires with a square cross section as a function of the lateral size $d_1=d_2=d$. Panels (a) and (b) show the results for the InSb nanowirezs and panels (c) and (d) show the results for the GaSb nanowires. The calculated data are presented by symbels ``$*$" and the solid lines are the results of fittings based on Eq.~(\ref{eq06}) with the fitting parameters listed in Table~\ref{tab:Table2}. The insets show the zoom-in plots of the calculated confinement energies in the nanowires at large sizes.}
      \label{Fig:[001]InSbconE}
  \end{figure}

\begin{figure}[t]
\centering
\includegraphics[width=85mm]{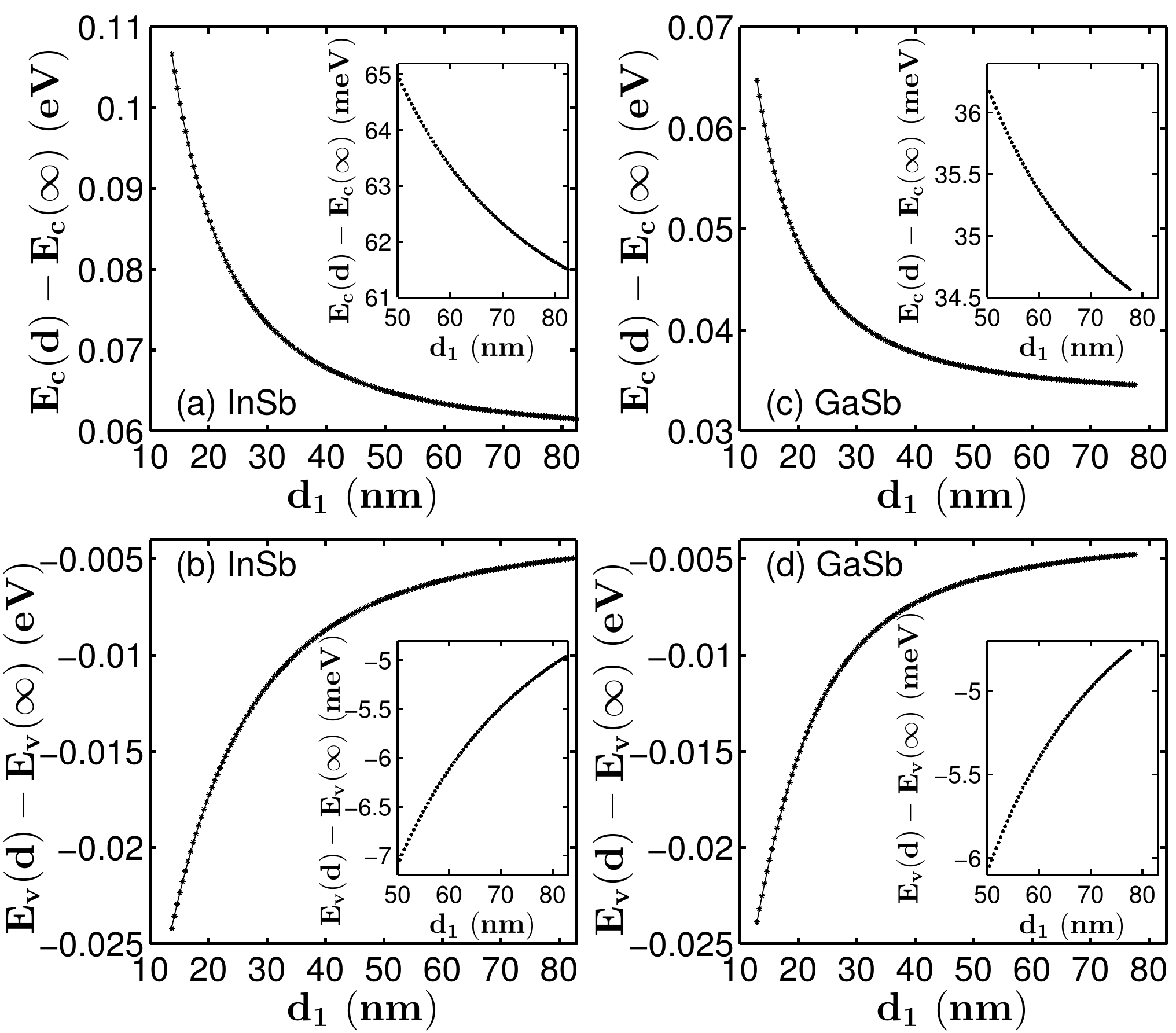}
      \caption{Lowest conduction band electron and highest valence band hole confinement energies in the [001]-oriented InSb and GaSb nanowires with a rectangular cross section. Panels (a) and (b) show the results for the InSb nanowires with fixed size $d_2=13.7$ nm and varying size $d_1=13.7$ nm to $82.5$ nm. Panels (c) and (d) show the results for the GaSb nanowires with fixed size $d_2=12.9$ nm and varying size $d_1=12.9$ nm to $77.6$ nm. The calculated data are presented by symbels ``$*$" and the solid lines are the results of fittings based on Eq.~(\ref{eq06}) with the fitting parameters listed in Table~\ref{tab:Table2}. The insets show the zoom-in plots of the calculated confinement energies in the nanowires at large sizes.}
      \label{Fig:[001]GaSbconE}
\end{figure}

In Fig.~\ref{Fig:[001]InSbbandsq} to Fig.~\ref{Fig:[001]GaSbbandrec}, it is seen that the lowest conduction and topmost valence band edges of the nanowires move apart in energy due to quantum confinement. For practical use, we fit the edge energies of the lowest conduction band and the topmost valence band, $E_{c}(d)$ and $E_{v}(d)$, of the [001]-oriented InSb nanowire and of the [001]-oriented GaSb nanowire to the following expression:
\begin{equation}
\label{eq06}
\Delta E_\alpha(d) =E_\alpha(d)-E_\alpha (\infty)=\frac{1}{p_{1} d^{2}+p_{2} d+p_{3}}+p_{4}
\end{equation}
where $p_{1}$, $p_{2}$, $p_{3}$, and $p_{4}$ are parameters to be determined by the fitting, $E_{\alpha}(\infty)=E_c$ or $E_v$ are the band edge energies of the relevant bulk materials taken from Ref.~\onlinecite{Klimech-1}, $\Delta E_\alpha(d)$ and $E_\alpha(d)$ are given in units of eV, and $d$ is taken in units of nm. Here we note that $p_{4}$ is nonzero only for the [001]-oriented nanowires with a rectangular cross section. The results of the fittings are  shown in Fig.~\ref{Fig:[001]InSbconE} for the nanowires with a square cross section and in Fig.~\ref{Fig:[001]GaSbconE} for the nanowires with a rectangular cross section with the fitting parameters listed in Table~\ref{tab:Table2}. Here we note that, as for the [001]-oriented InSb nanowires with a rectangular cross section,  we only present the results of the fittings for a fixed size $d_{2}$. For example, with the obtained fitting parameters listed in Table~\ref{tab:Table2}, Figs.~\ref{Fig:[001]GaSbconE}(a) and \ref{Fig:[001]GaSbconE}(b) show the results of the fittings for the [001]-oriented InSb nanowires with a rectangular cross section of  fixed size $d_{2}=13.7$ nm and varying size $d_1$ from 13.7 nm to 82.5 nm, while Figs.~\ref{Fig:[001]GaSbconE}(c) and \ref{Fig:[001]GaSbconE}(d) show the results of the fittings for the [001]-oriented GaSb nanowires with a rectangular cross section of fixed size $d_{2}=12.9$ nm and varying size $d_1$ from 12.9 nm to 77.6 nm.

Figures~\ref{Fig:[001]InSbconE} and \ref{Fig:[001]GaSbconE} show that the quantum confinement effect on the conduction band states of the InSb nanowires is very strong in consistence with the fact of small electron effective mass in the material. For example, the quantum confinement energy of electrons at the conduction band edge of the [001]-oriented InSb nanowire with a square cross section of size $d\sim 15$ nm is seen to be about 95 meV. Even when the InSb nanowire has a cross section size of $d\sim 55$ nm, the electron quantization energy at the conduction band edge can still be larger than 10 meV, as can be seen in the inset of Fig.~\ref{Fig:[001]InSbconE}(a). For the [001]-oriented GaSb nanowire with a square cross sectionof size $d\sim 15$ nm, the quantum confinement energy of electrons at the conduction band edge is about 51 meV, still large bu smaller than the corresponding value of the InSb nanowire with the same cross section size. In order to achieve the 95 meV quantum confinement energy of electrons at the conduction band edge, the GaSb nanowire cross section size needs to become smaller than 10 nm. In addition, it can be estimated that the quantum confinement energy of electrons at the conduction band edge of the GaSb nanowire becomes less than 5 meV when the nanowire cross section size becomes larger than 55 nm. The quantum confinement energies of holes in the valence bands of the [001]-oriented InSb and GaSb nanowires are generally very small when compared with the corresponding quantum confinement energies of electrons at the conduction band edge of the nanowires. For example, at the cross section size of $\sim 15$ nm, the quantum confinement energy of holes at the valence band edge of the [001]-oriented InSb (GaSb) nanowire with a square cross section is just about 20 meV (18 meV). Similar quantum confinement properties are also found in the [001]-oriented InSb and GaSb nanowires with a rectangular cross section. Nevertheless, it can be seen that at the limit of large values of size $d_1$, the quantum confinement energies of electrons and holes at the band edges of the nanowires with a rectangular cross section approach their respective constant values of $p_4$ listed in \ref{tab:Table2}. These values could be used to estimate quantum confinement energies of electron and holes in an InSb quantum well with width $d_2=13.7$ nm and in a GaSb quantum well with width $d_2=12.9$ nm.  

\begin{figure}[t]
\begin{center}
\includegraphics[width=85mm]{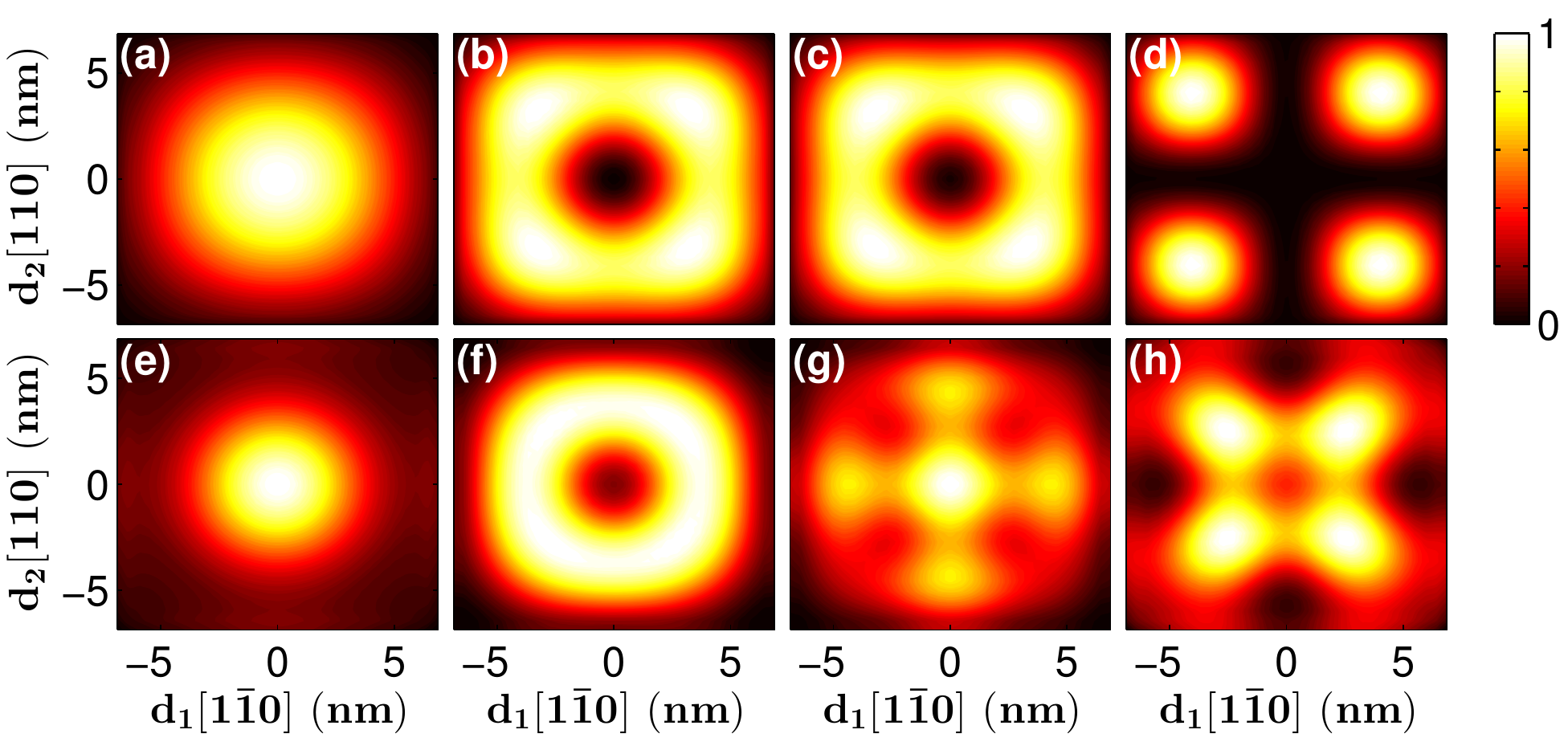}
\caption{(Color online) Wave functions of the four lowest conduction band states and the four highest valence band states at the $\Gamma$-point of the [001]-oriented InSb nanowire with a square cross section of size 13.7$\times$13.7 nm$^{2}$. The wave functions are presented by the probability distributions on a (001)-layer of cation In atoms whose value at each cation atomic site is calculated by summing up the squared amplitudes of all the atomic orbital components on the cation atomic site and is normalized within each panel by the highest value found in the panel. Panels (a) to (d) show the wave functions of the four lowest conduction band states at the $\Gamma$-point, while panels (e) to (h) show the wave functions of the four highest valence band states at the $\Gamma$-point.}
\label{Fig:InSb60to60wave}
\end{center}
\end{figure}

\begin{figure}[t]
\begin{center}
\includegraphics[width=85mm]{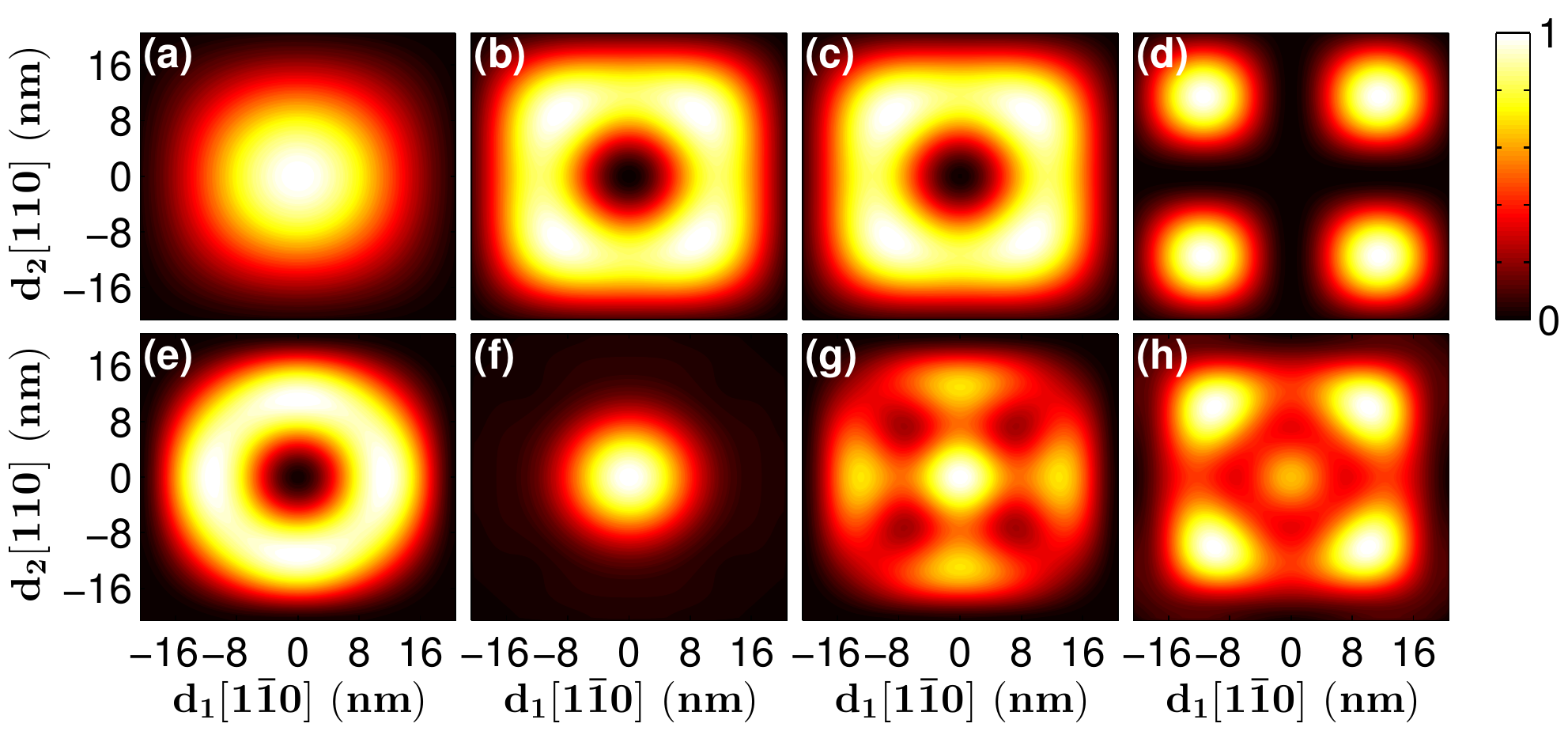}
\caption{(Color online) The same as in Fig.~\ref{Fig:InSb60to60wave} but for the [001]-oriented InSb nanowire with a square cross section of size 41.2$\times$41.2 nm$^{2}$.}
\label{Fig:InSb120to120wave}
\end{center}
\end{figure}

\subsection{Wave functions of [001]-oriented InSb and GaSb nanowires}

The wave functions of the band states of the [001]-oriented InSb and GaSb nanowires have been calculated. The representative results are shown in Fig.~8 to Fig.~13. In these figures, a wave function is represented by the probability distribution on a (001)-plane of cation (In or Ga) atoms with the probability at each atomic site calculated by summing up the squared amplitudes of all the atomic orbital components on that site and scaled against the maximum value with each graph. Also, we note that only the wave function of one of two spin-degenerate states is presented because the other one has an identical spacial distribution.

Figure~\ref{Fig:InSb60to60wave} shows the calculated wave functions of the four lowest conduction band states and the four highest valence band states at the  $\Gamma$-point for the [001]-oriented InSb nanowire with a square cross section of size 13.7$\times$13.7 nm$^{2}$.
It is seen in the figure that the wave functions of the lowest and the fourth lowest conduction band states at the $\Gamma$-point have the propability distributions similar to what one would obtained from a spin-orbit decoupled, single-band effective mass theory. However, the wave functions of the second and third lowest conduction band states at the $\Gamma$-point show the propability distributions of donut shape which are very different from what one would obtained from a spin-orbit decoupled, single-band effective mass theory. Thus, in the study of the electronic properties of these nanowires, it is important to include spin-orbit interaction. Another feature found in the calculations for the second and third lowest conduction band states at the $\Gamma$ -point is that the spacial distributions of the wave functions of the two states look almost identical, which also could not be obtained from a spin-orbit decoupled, single-band effective mass theory but is consistent with the fact that the two band states are nearly degenerate in energy. The highest valence band state of the [001]-oriented InSb nanowire with a square cross section of size 13.7$\times$13.7 nm$^{2}$ at the $\Gamma$-point has a similar propability distribution as the lowest conduction band of the nanowire, but is more localized to the center of the nanowire. The second highest valence band state of the nanowire at the $\Gamma$-point also has a similar propability distribution as the second or third lowest conduction band state of the nanowire at the $\Gamma$-point, but is slightly more localized to the center of the nanowire. In contrast, the third and the fourth highest valence band states at the $\Gamma$-point are very different from the corresponding third and fourth lowest conduction states. These states show complex but four-fold symmetric probability distribution pattens and are also more localized to the inner region of the nanowire.

Figure~\ref{Fig:InSb120to120wave} shows the calculated wave functions of the four lowest conduction band states and the four highest valence band states at the  $\Gamma$-point for the [001]-oriented InSb nanowire with a square cross section of the size $41.2\times 41.2$ nm$^{2}$. When compared with the probability distributions shown in Fig.~\ref{Fig:InSb60to60wave}, it is seen that as the size of the nanowire increases, the  wave functions of the four lowest conduction band states at the $\Gamma$-point do not show significant changes in the probability distributions. The same is also found for the third highest valence band state of the nanowire at the $\Gamma$-point. However, the wave functions of the first two highest valence band states exchange their probability distribution patterns as the size of the nanowire increases. The result is consistent with the symmetry properties of the two states as shown in Figs.~\ref{Fig:[001]InSbbandsq}. For the nanowire with the cross section of size $13.7\times 13.7$ nm$^2$, the highest and the second highest valence band state at the $\Gamma$-point are $\Gamma_7$- and $\Gamma_6$-symmetric, while for the nanowire with the cross section of size $41.2\times 41.2$ nm$^2$ the two corresponding valence band states are $\Gamma_6$- and $\Gamma_7$-symmetric, i.e., the two states have reversed their symmetry ordering. The fourth highest valence band state shows a significant change in the probability distribution of the wave function as the size of the nanowire increases--it has a large probability distribution at the center of the nanowire when the nanowire has a size of $41.2\times 41.2$ nm$^2$--although its four-fold symmetric property remains unchanged. This change in the probability distribution arises from the fact that the wave function of the fourth highest valence band state has been mixed into with the characteristics of the wave function of the fifth highest valence band state, since as can be seen in Fig.~\ref{Fig:[001]InSbbandsq} the two states at the nanowire with a large cross section size move closer in energy  and have the same symmetry.

\begin{figure}[t]
\begin{center}
\includegraphics[width=85mm]{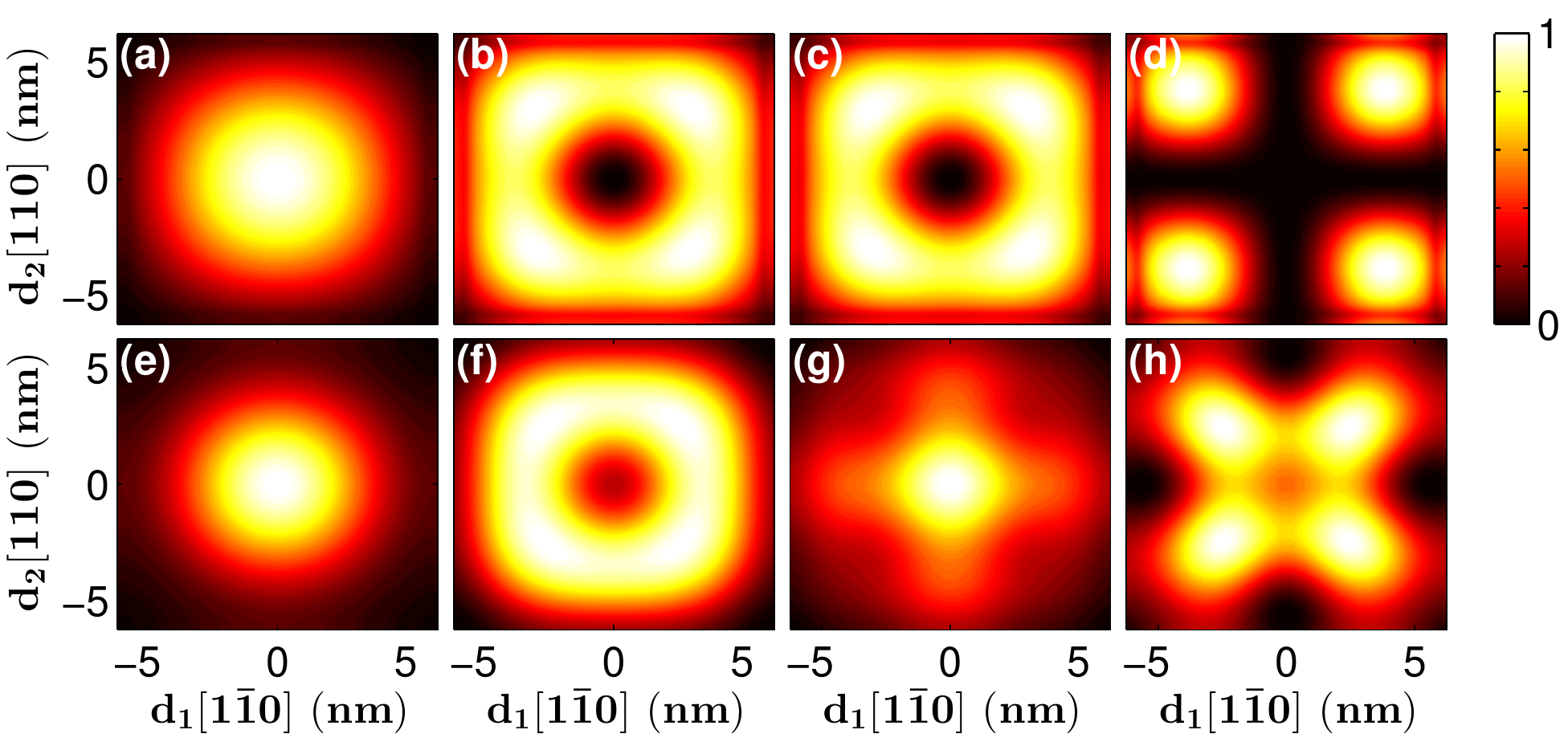}
\caption{(Color online) Wave functions of the four lowest conduction band states and the four highest valence band states at the $\Gamma$-point of the [001]-oriented GaSb nanowire with a square cross section of size 12.9$\times$12.9 nm$^{2}$. The wave functions are presented by the probability distributions on a (001)-layer of cation Ga atoms whose value at each cation atomic site is calculated by summing up the squared amplitudes of all the atomic orbital components on the cation atomic site and is normalized within each panel by the highest value found in the panel. Panels (a) to (d) show the wave functions of the four lowest conduction band states at the $\Gamma$-point, while panels (e) to (h) show the wave functions of the four highest valence band states at the $\Gamma$-point.}
\label{Fig:GaSb120to120wave}
\end{center}
\end{figure}

\begin{figure}[t]
\begin{center}
\includegraphics[width=85mm]{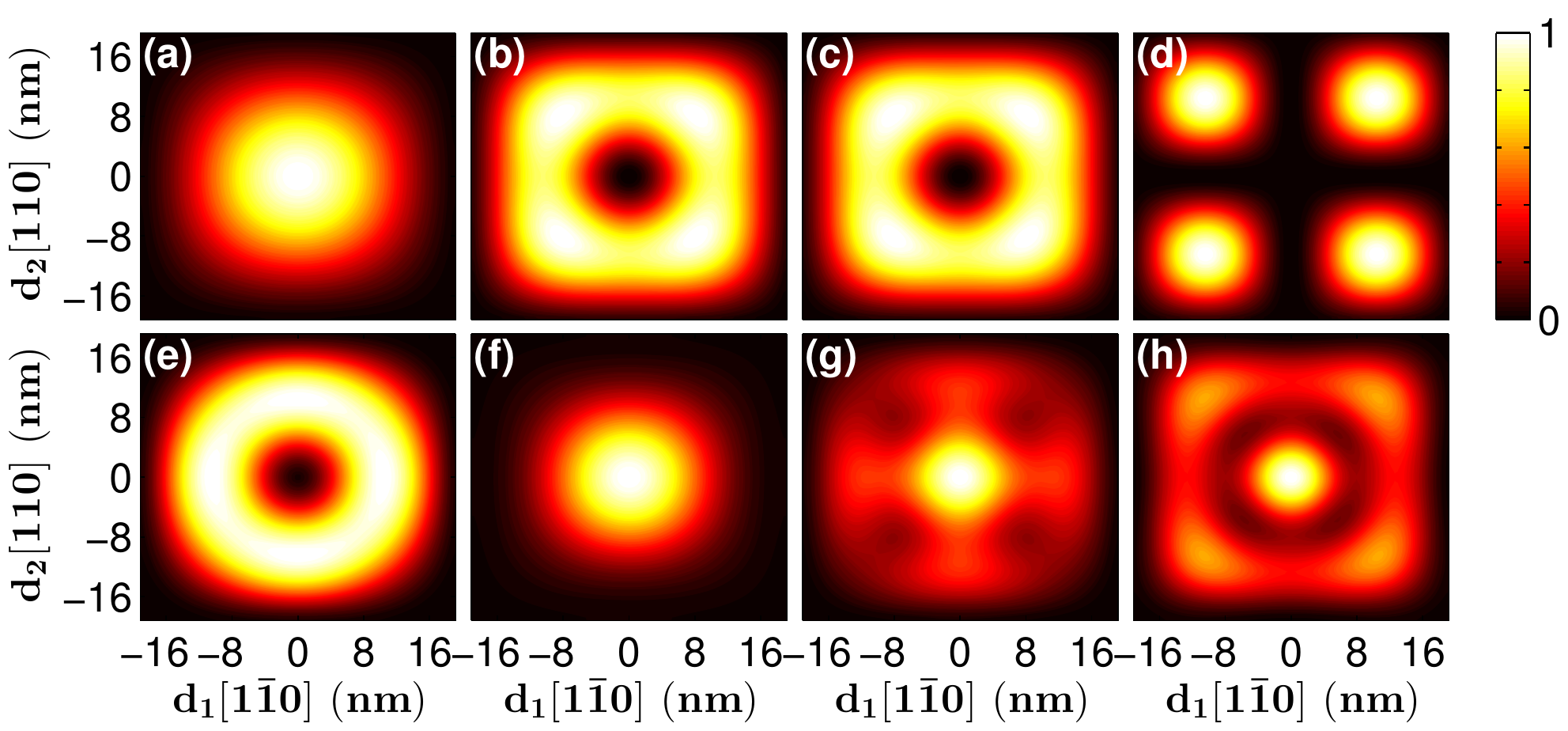}
\caption{(Color online) The same as in Fig.~\ref{Fig:GaSb120to120wave} but for the [001]-oriented GaSb nanowire with a square cross section of size 38.8$\times$38.8 nm$^{2}$.}
\label{Fig:GaSb180to180wave}
\end{center}
\end{figure}

Figures~\ref{Fig:GaSb120to120wave} and Fig.~\ref{Fig:GaSb180to180wave} show the calculated wave functions for the band states of the [001]-oriented GaSb nanowire with a square cross section of different sizes. Figure~\ref{Fig:GaSb120to120wave} shows the wave functions of the four lowest conduction band states and the four highest valence band states at the $\Gamma$-point for the nanowire with the cross section size of $12.9\times 12.9$ nm$^2$. It is seen that the wave functions of these band states of the GaSb nanowire resemble well their corresponding states in the [001]-oriented InSb nanowire with a small cross section size of $13.7\times 13.7$ nm$^2$ shown in Fig.~\ref{Fig:InSb60to60wave}. Figure~\ref{Fig:GaSb180to180wave} shows the wave functions of the four lowest conduction band states and the four highest valence band states at the $\Gamma$-point for the nanowire with the cross section size of $38.8\times 38.8$ nm$^2$. As in the case for the [001]-oriented InSb nanowire, the probability distribution patterns of the wave functions of the four lowest conduction band states at the $\Gamma$-point remain almost unchanged as the cross section size of the [001]-oriented GaSb nanowire is increased. Also as in the case for the [001]-oriented InSb nanowire, with increasing the cross section size of the [001]-oriented GaSn nanowire, the probability distribution patterns of the two highest valence band states at the $\Gamma$-point are inter-changed and the fourth highest valence state shows mixed-in characteristics of the fifth highest valence band state. All these results are again consistent with symmetry properties of the states as shown in Fig.~\ref{Fig:[001]GaSbsq}.   

\begin{figure}[t]
\begin{center}
\includegraphics[width=85mm]{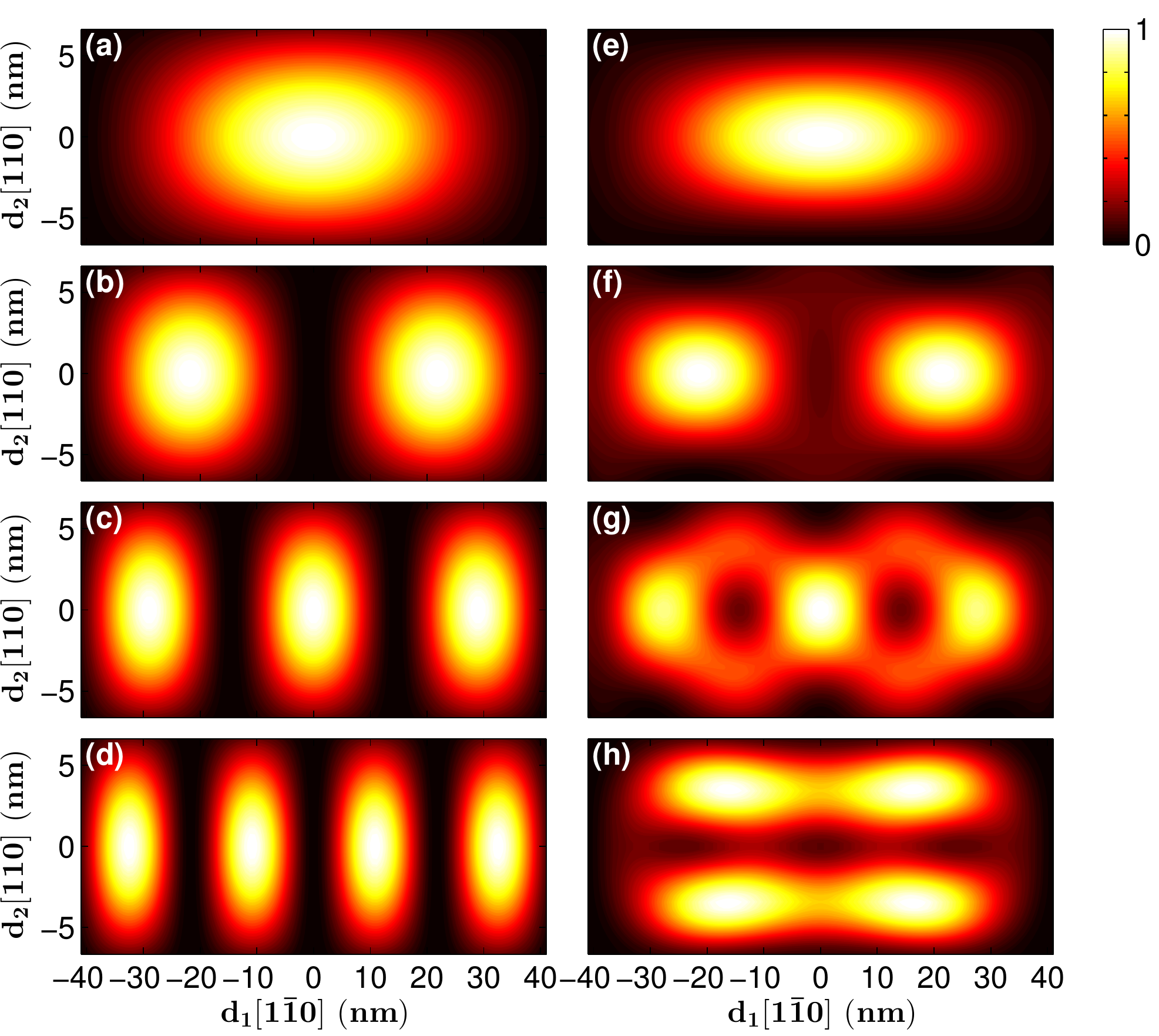}
\caption{(Color online) Wave functions of the four lowest conduction band states and the four highest valence band states at the $\Gamma$-point of the [001]-oriented InSb nanowire with a rectangular cross section of size 82.5$\times$13.7 nm$^{2}$. The wave functions are presented by the probability distributions on a (001)-layer of cation In atoms whose value at each cation atomic site is calculated by summing up the squared amplitudes of all the atomic orbital components on the cation atomic site and is normalized within each panel by the highest value found in the panel. Panels (a) to (d) show the wave functions of the four lowest conduction band states at the $\Gamma$-point, while panels (e) to (h) show the wave functions of the four highest valence band states at the $\Gamma$-point.}
\label{Fig:InSb360to60wave}
\end{center}
\end{figure}

\begin{figure}[t]
\begin{center}
\includegraphics[width=85mm]{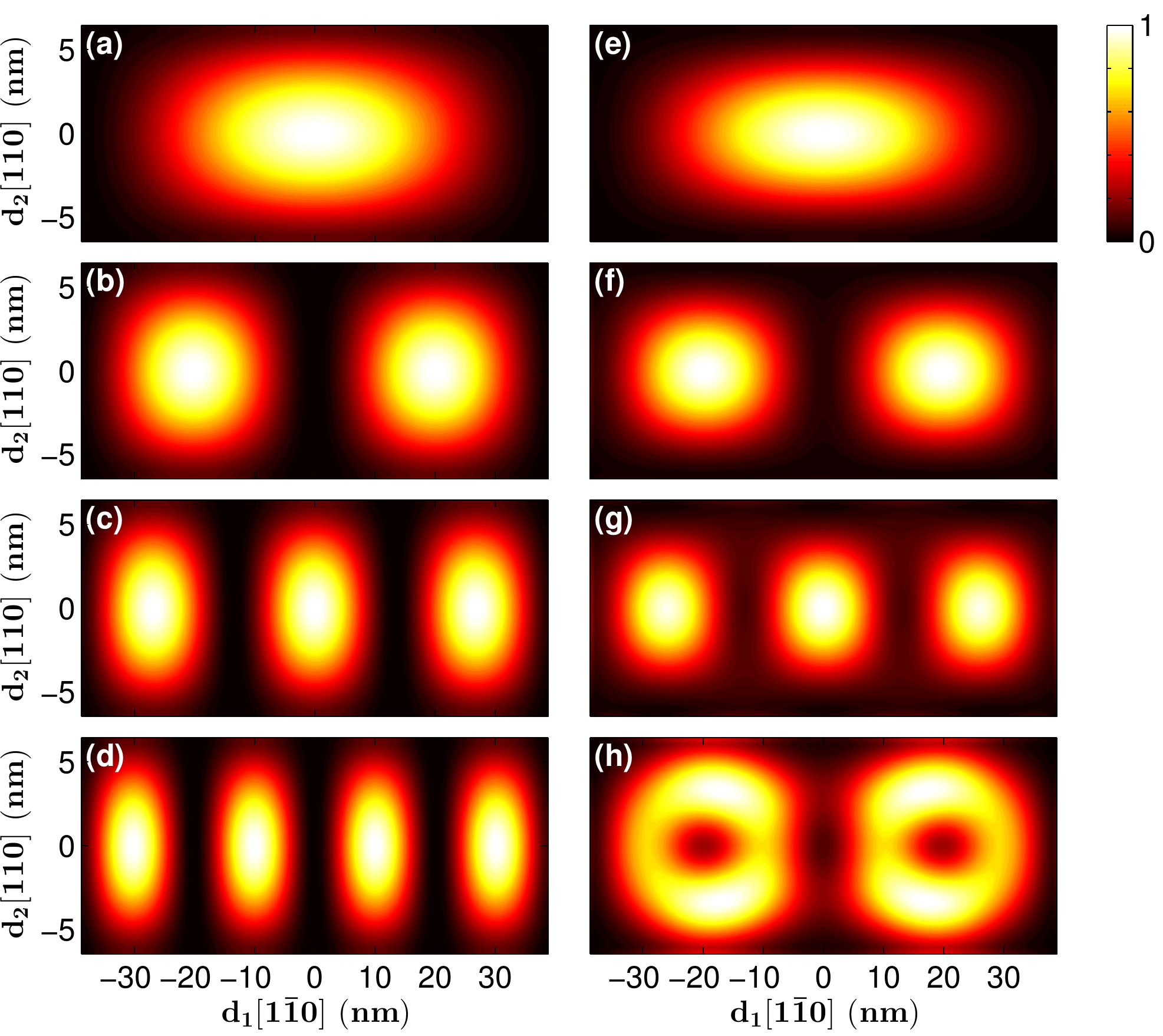}
\caption{(Color online) The same as in Fig.~\ref{Fig:InSb360to60wave} but for the [001]-oriented GaSb nanowire with a rectangular cross section of size 77.6$\times$12.9 nm$^{2}$. }
\label{Fig:GaSb360to60wave}
\end{center}
\end{figure}

\begin{figure*}[t]
 \centering
      \includegraphics[width=7in]{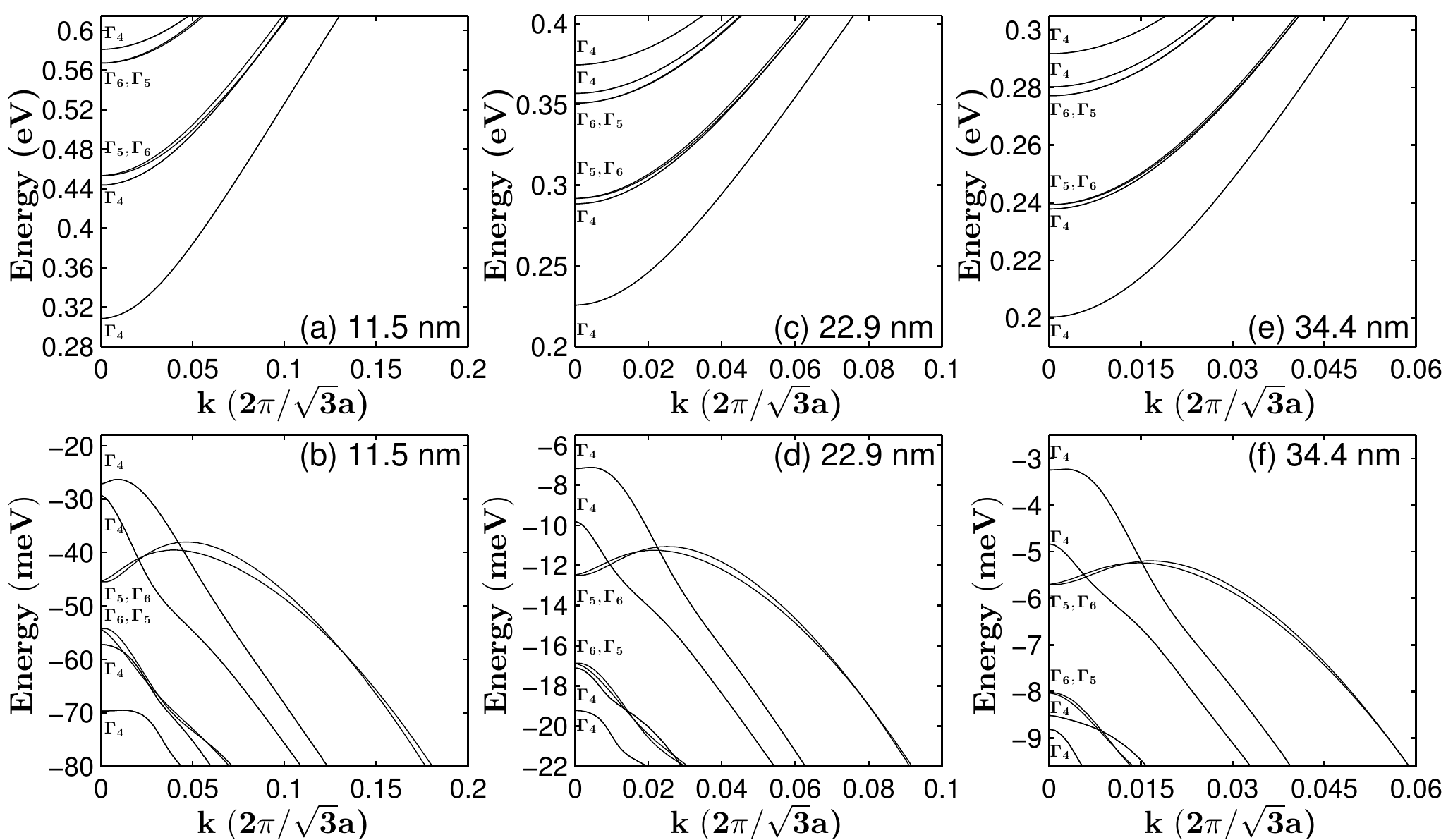}
       \caption{Band structures of [111]-oriented InSb nanowires with a hexagonal cross section of the lateral sizes of (a) and (b) $d$=11.5 nm, (c) and (d) $d$=22.9 nm, and (e) and (f) $d$=34.4 nm. The symmetries of the band states are marked according to the irreducible representations, $\Gamma_{4}$, $\Gamma_{5}$, and $\Gamma_{6}$ of the $C_{3v}$ double point group. All the $\Gamma_4$ bands are double degenerate, while $\Gamma_5$ and $\Gamma_6$ bands are no-degenerate except for at the $\Gamma$-point at which a $\Gamma_5$ band state is degenerate with a $\Gamma_6$ band state due to the Kramers time reversal symmetry. In the figure, the $\Gamma_5$ and $\Gamma_6$ bands are labeled in such a way that the first (second) symmetry symbol corresponds to a band that initially has a lower (higher) energy after splitting as the wave vector moves away from the $\Gamma$-point. }
        \label{Fig:[111]InSbband}
\end{figure*}

\begin{figure*}[t]
 \centering
      \includegraphics[width=7in]{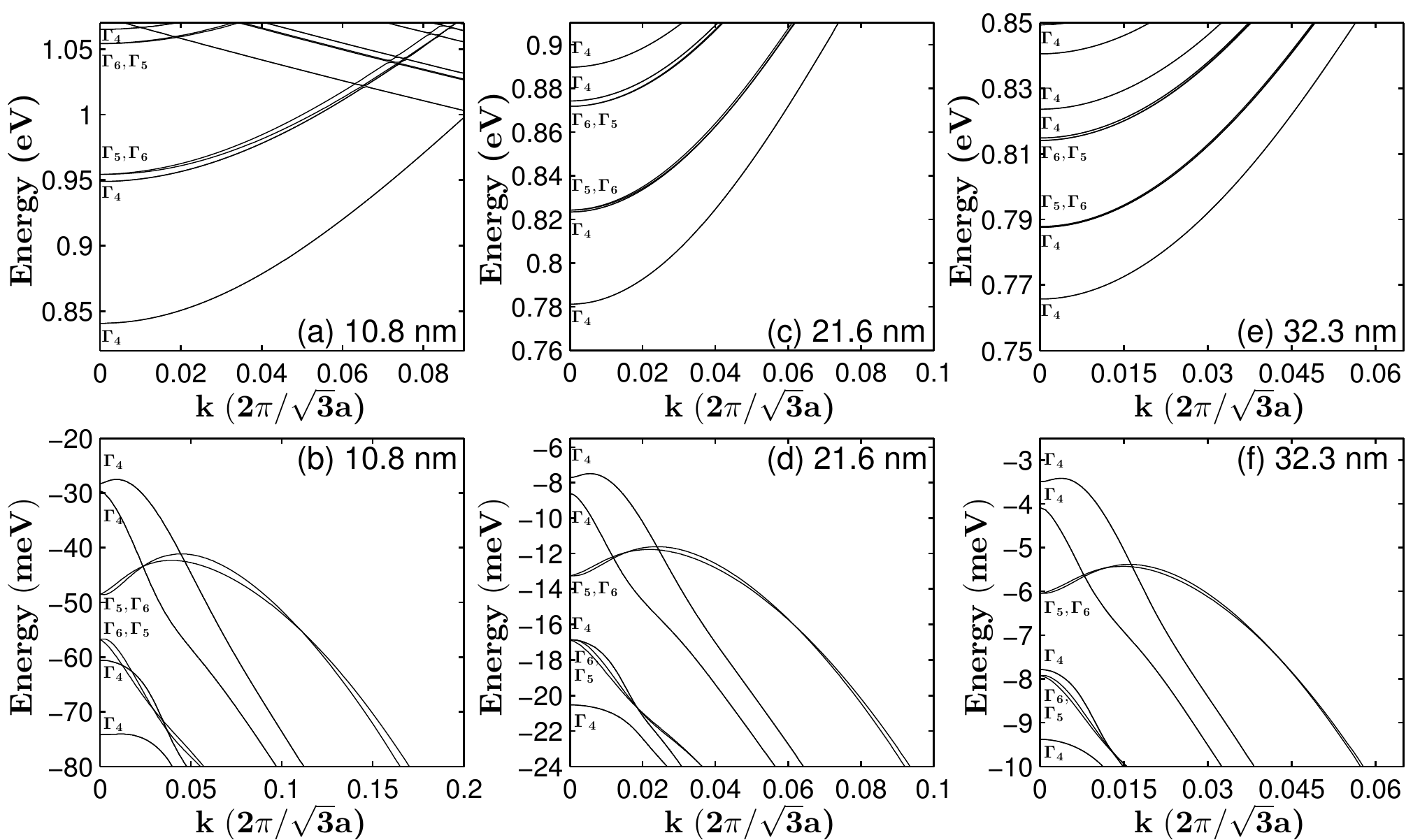}
       \caption{Band structures of [111]-oriented GaSb nanowires with a hexagonal cross section of the lateral sizes of (a) and (b) $d$=10.8 nm, (c) and (d) $d$=21.6 nm, and (e) and (f) $d$=32.3 nm. The symmetries of the band states are marked according to the irreducible representations, $\Gamma_{4}$, $\Gamma_{5}$, and $\Gamma_{6}$ of the $C_{3v}$ double point group. All the $\Gamma_4$ bands are double degenerate, while $\Gamma_5$ and $\Gamma_6$ bands are non-degenerate except for at the $\Gamma$-point at which a $\Gamma_5$ band state is degenerate with a $\Gamma_6$ band state due to the Kramers time-reversal symmetry. In the figure, the $\Gamma_5$ and $\Gamma_6$ bands are labeled in such a way that the first (second) symmetry symbol corresponds to a band that initially has a lower (higher) energy after splitting as the wave vector moves away from the $\Gamma$-point.}
        \label{Fig:[111]GaSbband}
\end{figure*}
Figures~\ref{Fig:InSb360to60wave} and Fig.~\ref{Fig:GaSb360to60wave} show the wave functions of the four lowest conduction band and four highest valence band states at $\Gamma$-point of the [001]-oriented InSb nanowire with a rectangular cross section of size $82.5\times 13.7$ nm$^2$ and the wave function of these band states of the [001]-oriented GaSb nanowire with a rectangular cross section of size $77.6\times 12.9$ nm$^2$, respectively. It is seen that the lowest conduction band states in both nanowires show regular probability distribution patterns with one, two, three and four maxima present in the wave functions of the first, second, third, and fourth lowest conduction band state, respectively. These results are similar to what one would obtain from simple, spin-orbit decoupled, single-band effective mass theory. However, not all top valence band states resemble the characteristics of the wave functions predicted based on a simple, spin-orbit decoupled, single-band effective mass theory. For the [001]-oriented InSb nanowire with a rectangular cross section of the size $82.5\times 13.7$ nm$^2$, only the top two valence band states show the characteristics of the wave functions predicted by a simple effective mass theory. The third and fourth highest valence band states show very different characteristics from what would be predicted using a sipmle, spin-orbit decoupled, single-band effective mass model. The third highest valence band state is shaped as an opened peanut shell and the fourth highest valence band state is shaped as a pair of parallel dumbbells. These results are consistent with the valence band structure shown in Fig.~\ref{Fig:[001]InSbbandrec}, where it is seen that only the top two valence bands show good parabolic dispersions around the $\Gamma$-point. For the [001]-oriented GaSb nanowire with a rectangular cross section of size $77.6\times 12.9$ nm$^2$, the three highest valence band states show the results one would obtain from a simple, spin-orbit decoupled, single-band effective mass model, in consistence with the fact that the three highest valence bands show parabolic dispersions around the $\Gamma$-point as seen in Fig.~\ref{Fig:[001]GaSbbandrec}, while the wave function of the fourth highest valence band state has a double eye shape and does not resemble the wave function predicted based on a simple, spin-orbit decoupled, single-band effective mass model. 

\section{Electronic Structures of [111]-oriented InSb and GaSb nanowires}

Semiconductor InSb and GaSb nanowires grown along the [111] crystographic direction often have a hexagonal cross section and $\{1\bar{1}0\}$ facets. The lateral size of these nanowires can be defined as the distance $d$ between two parallel surfaces on each nanowire. In our atomistic tight-binding model, the lateral size takes discrete values of $d=Ia/\sqrt{2}$, where $a$ is the bulk lattice constant and $I$ is a positive interger number. The [111]-oriented InSb and GaSb nanowires with a hexagonal cross section are symmetric under the operations of the $C_{3v}$  point group whose corresponding double point group has one two-dimensional irreducible representations $\Gamma_{4}$ and two one-dimensional irreducible representations $\Gamma_{5}$ and $\Gamma_{6}$.\cite{Melvin-1} However, the characters of the two one-dimensional irreducible representations  $\Gamma_{5}$  and $\Gamma_{6}$ are two complex conjugate, imaginary numbers. Thus, at the $\Gamma$-point, where the band structure Hamiltonians of the nanowires are symmetric under time reversal, the $\Gamma_5$ and $\Gamma_6$ band states are degerated, forming Kramers doublets in the energy spectra. These doubletes together with the double degenerate $\Gamma_4$ band states ensure that all the band states at the $\Gamma$-point are double degenerate states. At a finite wave vector point, the band structure Hamiltonian does not possess the time reversal symmetry and the Kramers degenerate $\Gamma_5$ and $\Gamma_6$ states would split into non-degenerate states. However, the $\Gamma_4$ bands will remain double degenerate at all wave vector points.

\subsection{Band structures of [111]-oriented InSb and GaSb nanowires}

We first present the calculated band structures of the [111]-oriented InSb and GaSb nanowires. Figures~\ref{Fig:[111]InSbband} shows the calculated band structures of the [111]-oriented InSb nanowires with a hexagonal cross section of sizes $d=11.5$, 22.9, and 34.4 nm. Figure~\ref{Fig:[111]GaSbband} shows the calculated band structures of the [111]-oriented  GaSb nanowires with a hexagonal cross section of sizes $d=10.8$, 21.6, and 32.3 nm. In general, as expected and as already seen in the band structures of the [001]-oriented InSb and GaSb nanowires, the band gaps of the [111]-oriented InSb and GaSb nanowires are both increased as the lateral size $d$ is decreased. Also, as seen in the band structures of the [001]-oriented InSb and GaSb nanowires, the conduction bands of the [111]-oriented InSb and GaSb nanowires near the band gaps are simple, parabolic bands. However, in contrast, the top valence bands of the [111]-oriented InSb and GaSb nanowires show distinct, different characteristics from the [001]-oriented InSb and GaSb nanowires.

\begin{figure}[t]
\begin{center}
\includegraphics[width=85mm]{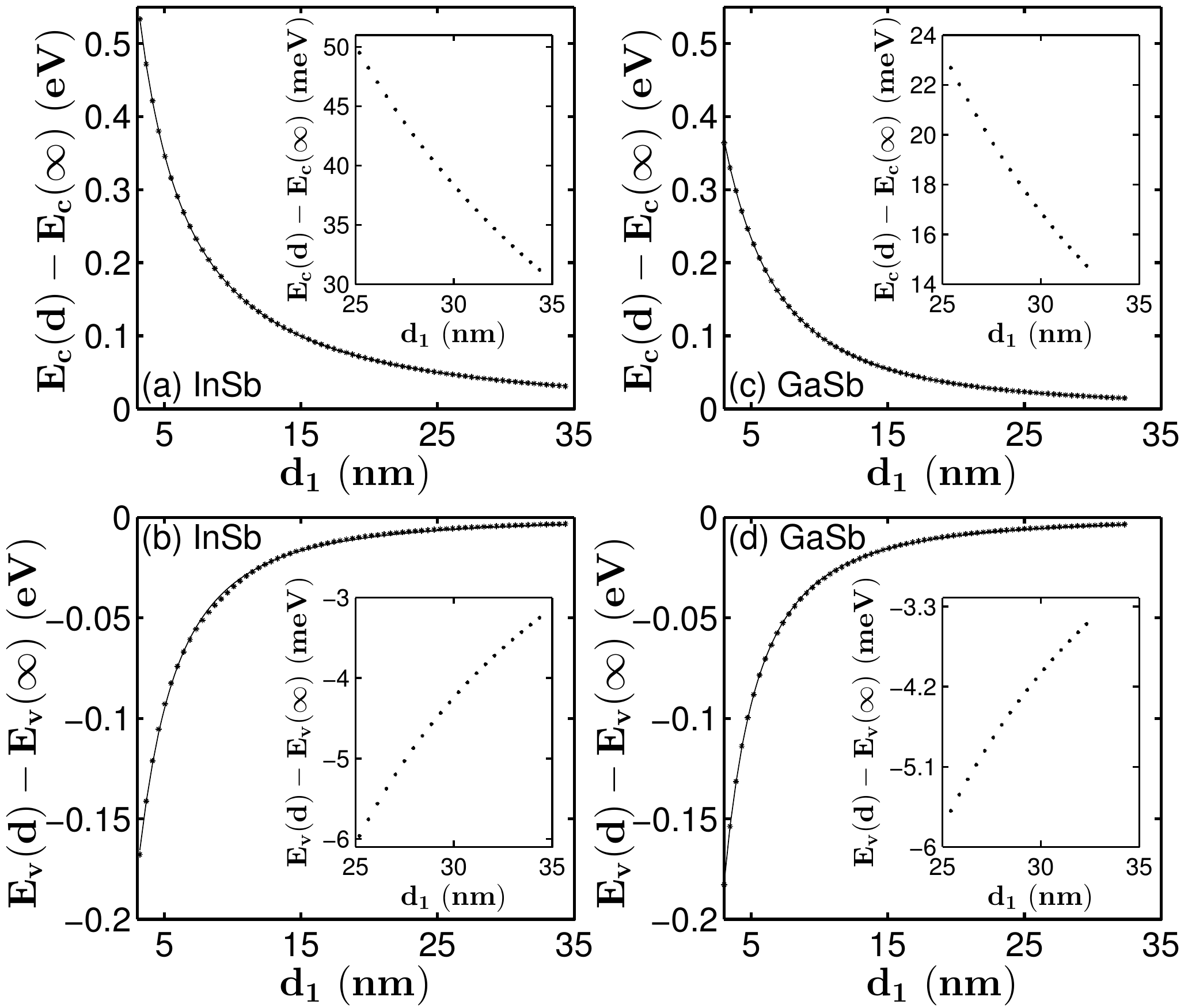}
\caption{Lowest conduction band electron and highest valence band hole confinement energies in the [111]-oriented InSb nand GaSb nanowires with a hexagonal cross section as a function of the cross section size $d$. Panels (a) and (b) show the results for the InSb nanowire and panels (c) and (d) show the results for the GaSb nanowire. The calculated data are presented by symbels ``$*$" and the solid lines are the results of fittings based on Eq.~(\ref{eq06}) with the fitting parameters listed in Table~\ref{tab:Table3}. The insets show the zoom-in plots of the calculated confinement energies in the nanowires at large sizes.}
\label{Fig:InGaSb111conE}
\end{center}
\end{figure}

In details, Figs.~\ref{Fig:[111]InSbband}(a), \ref{Fig:[111]InSbband}(c), and \ref{Fig:[111]InSbband}(e) show that the lowest conduction band of a [111]-oriented InSb nanowire with a hexagonal cross section is $\Gamma_4$ symmetruc and double degenerate. The second lowest conduction band is also $\Gamma_4$ symmetric and double degenerate. However, this band is very close to the next lowest $\Gamma_5$ and $\Gamma_6$ bands in energy, leading to the formation of a nearly four-fold degenerate band. The next two lowest conduction bands are $\Gamma_5$ and $\Gamma_6$ bands. These two bands are close to the next $\Gamma_4$ bands in energy, forming another nearly four-fold degenerate conduction band. Figures~\ref{Fig:[111]InSbband}(b), \ref{Fig:[111]InSbband}(d), and \ref{Fig:[111]InSbband}(f) show that the two highest valence bands of a [111]-oriented InSb nanowire with a haxongal cross section are $\Gamma_4$ symmetric, double degenerate bands. The next two highest valence bands are a $\Gamma_5$ and a $\Gamma_6$ band whose states are, in general, very close in energy but non-degenerate, except for at the $\Gamma$-point at which the two bands are degenerate due to the time reversal symmetry. Also, as seen in Figs.~\ref{Fig:[111]InSbband}(b), \ref{Fig:[111]InSbband}(d), and \ref{Fig:[111]InSbband}(f), with the wave vector moves away from the $\Gamma$-point, the $\Gamma_5$ and $\Gamma_6$ bands increase in energy and will cross the two highest $\Gamma_4$ bands. However, no pronounced double maximum structure is formed in the valence band structure of the [111]-oriented InSb nanowire, in difference from the valence band structure of a [001]-oriented InSb nanowire with a square cross section [cf. Fig.~\ref{Fig:[001]InSbbandsq}]. 
The next two highest valence bands of the [111]-oriented InSb nanowire are again the $\Gamma_5$ and $\Gamma_6$ bands with nearly degenerate energies (except for at the $\Gamma$-point where the two band states are precisely degenerate). These two bands are also close to the next highest $\Gamma_4$ band in energy, forming a four-fold nearly degenerate valence band. Figure~\ref{Fig:[111]GaSbband} shows that the band structure of a [111]-oriented GaSb nanowire exhibits similar characteristics as the [111]-oriented InSb nanowire, except for that in the conduction band structure of  the GaSb nanowire with a small size of $d=10.8$ nm shown in Fig.~\ref{Fig:[111]GaSbband}(a) a few folded bands are visible at high energies and that the band states in the four-fold nearly degenerate valence bands may have different symmetry orderings as seen in Figs.~\ref{Fig:[111]GaSbband}(b), \ref{Fig:[111]GaSbband}(d) and \ref{Fig:[111]GaSbband}(f). 

\begin{table*}[t]
\caption{Parameters $p_{1}$, $p_{2}$, and $p_{3}$ in Eq.~(\ref{eq06}) ontained by fitting the equation to the calculated energies at the $\Gamma$-point of the lowest conductance band and the highest valence band of the [111]-oriented InSb and GaSb nanowires with a hexagonal cross section (labeled by subscript ``{\em hex}"). Note that $p_4=0$ in Eq.~(\ref{eq06}) for the [111]-oriented nanowires with a hexagonal cross section.}
\begin{center}
\begin{tabular*}{120mm}{@{\extracolsep{\fill}}c c c c c c }
\hline\hline
Material & Nanowire & Band shift & $p_{1}$             &$p_{2}$             &$p_{3}$     \\
         &  type    &    (eV)    &(eV$^{-1}$nm$^{-2}$) &(eV$^{-1}$nm$^{-1}$)&(eV$^{-1}$) \\
\hline
InSb   &$[111]_{hex}$              &$\Delta E_{c}$  &0.01377  &0.44126  &0.31880 \\
\qquad    &$[111]_{hex}$       &$\Delta E_{v}$  &-0.20715  &-0.78595  &-1.38203 \\
\hline
GaSb   &$[111]_{hex}$       &$\Delta E_{c}$  &0.05150  &0.37413  &1.12596 \\
\qquad    &$[111]_{hex}$       &$\Delta E_{v}$  &-0.22274  &-0.86724  &-0.85199 \\
    \hline \hline
\end{tabular*}
\end{center}
\label{tab:Table3}
\end{table*}

As we mentioned above and as we already showed for the [001]-oriented InSb and GaSb nanowires, the conduction bands and the valence bands move apart in energy as the cross section sizes of the [111]-oriented InSb and GaSb nanowires are decreased due to quantum confinement. Again, for practical use, we fit the calculated edge energies of the conduction and valence bands, $E_{c}(d)$ and $E_{v}(d)$, of the [111]-oriented[ InSb and GaSb nanowires with a hexagonal cross section to Eq.~(\ref{eq06}). The results are shown in Fig.~\ref{Fig:InGaSb111conE} with the obtained fitting parameters listed in Table~\ref{tab:Table3}. Here, we note again that in difference from the nanowires with a rectangular cross section, parameter $p_4$ in Eq.~(\ref{eq06}) has been set to zero in the fittings for the band edges of the [111]-oriented InSb and GaSb nanowires. Again, it is seen that the quantum confinement is stronger for electrons in the conduction bands of an InSb nanowire than a GaSb nanowire with the same cross section size. For example, as can be extracted from the figure, the quantum confinement energy of electrons at the conduction band edge of the [111]-oriented InSb nanowire with a hexagonal cross section of size $d\sim 15$ nm is $\sim$100 meV, while the corresponding value for the [111]-oriented GaSb nanowire with a hexagonal cross section of size $d\sim 15$ nm is $\sim 55$ meV. The quantum confinement energies of holes in the valence bands are, in general, again comparably small in the [111]-oriented InSb and GaSb nanowires with hexagonal cross sections. These quantum confinement energies are only $\sim 16$ meV at the valence band edges of the [111]-oriented InSb and GaSb nanowires with a hexagonal cross section of size $d\sim 15$ nm, and become $\sim 3$ meV when the cross section size increases to $\sim 35$ nm. 

\subsection{Wave functions of [111]-oriented InSb and GaSb nanowires}

\begin{figure}[t]
\begin{center}
\includegraphics[width=85mm]{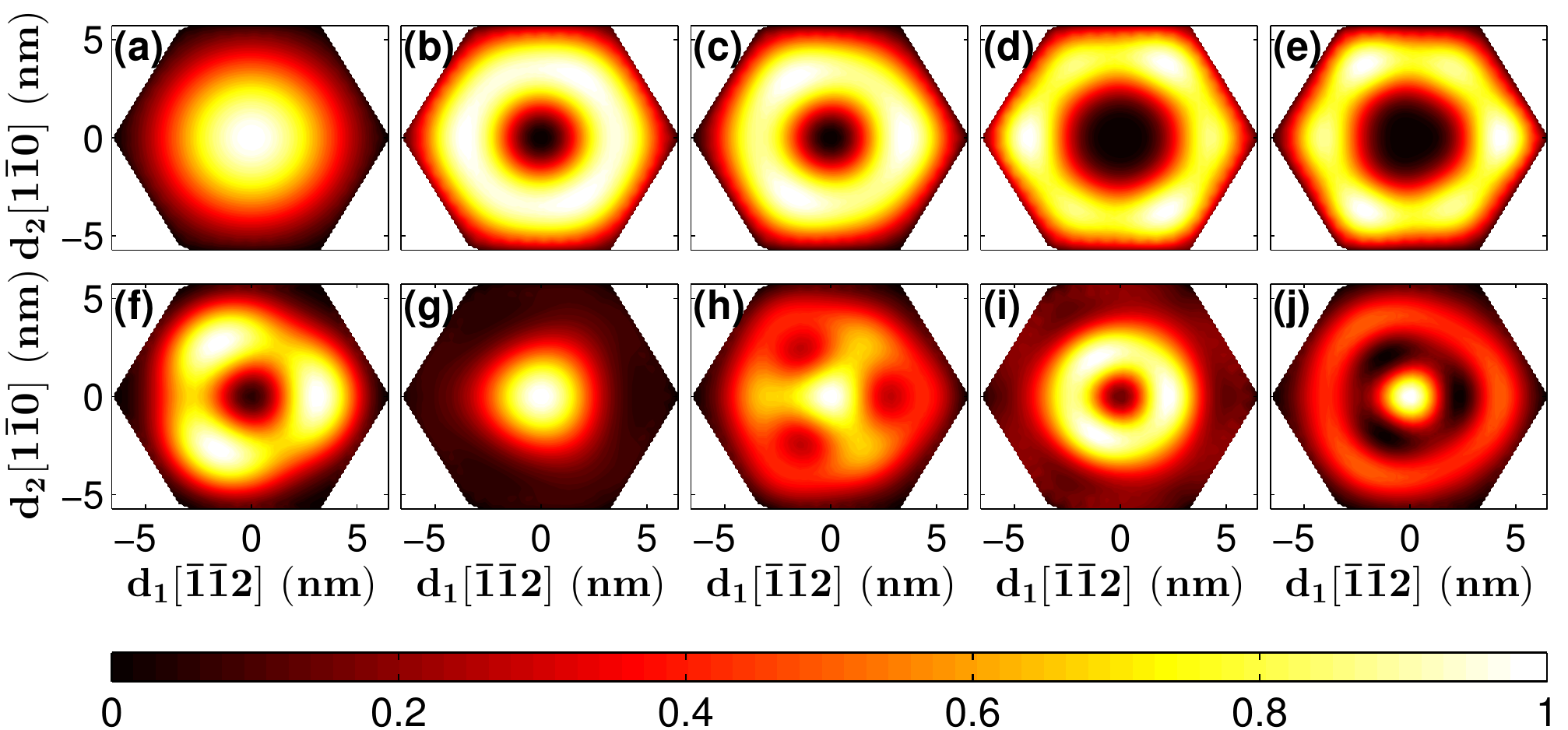}
\caption{(Color online) Wave functions of the five lowest conduction band states and the five highest valence band states at the $\Gamma$-point of the [111]-oriented InSb nanowire with a hexagonal cross section of size $d=11.5$ nm. The wave functions are presented by the probability distributions on a (111)-plane of cation In atoms whose value at each cation atomic site is calculated by summing up the squared amplitudes of all the atomic orbital components on the cation atomic site and is normalized within each panel by the highest value found in the panel. Panels (a) to (e) show the wave functions of the five lowest conduction band states at the $\Gamma$-point, while panels (f) to (j) show the wave functions of the five highest valence band states at the $\Gamma$-point.}
 \label{Fig:InSbN25}
\end{center}
\end{figure}

\begin{figure}[t]
\begin{center}
\includegraphics[width=85mm]{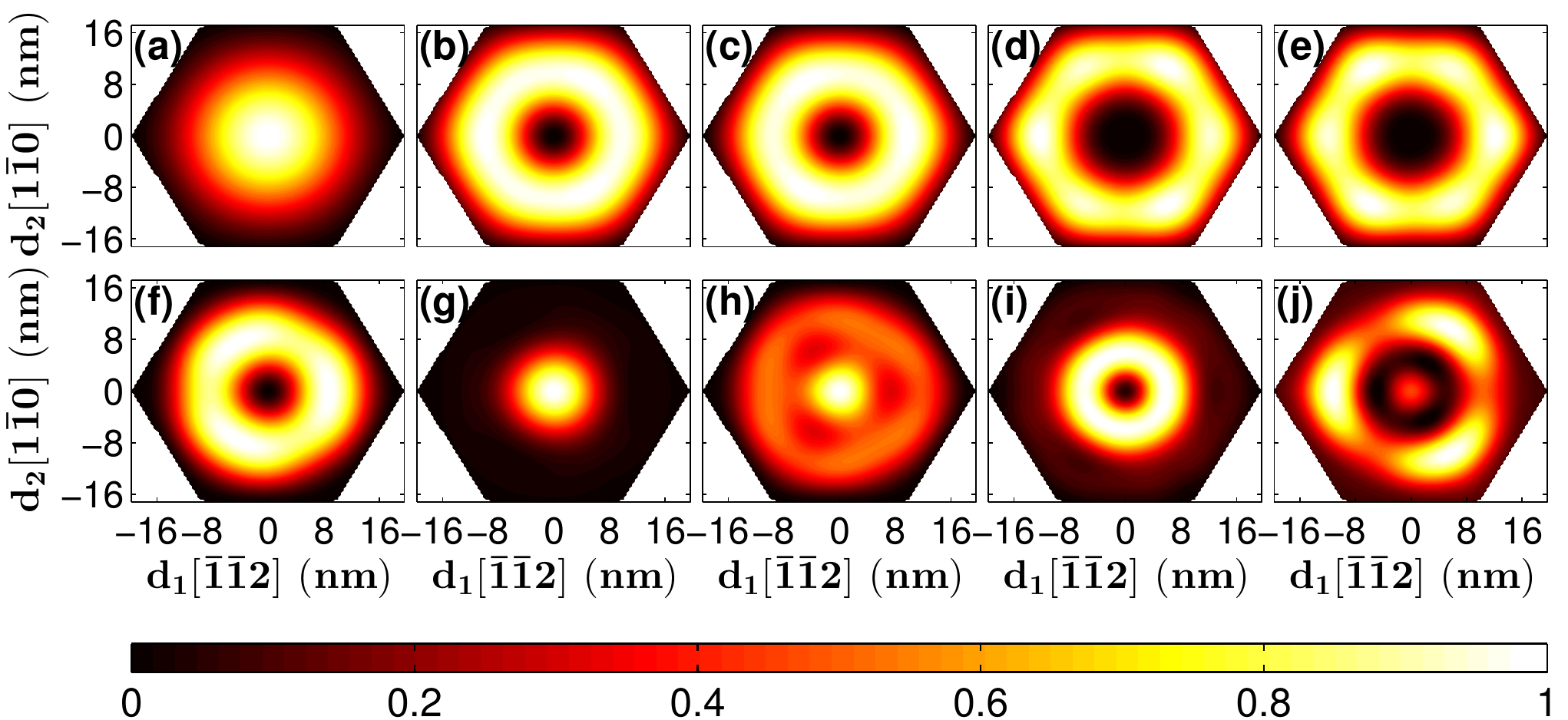}
\caption{(Color online) The same as in Fig.~\ref{Fig:InSbN25} but for the [111]-oriented InSb nanowire with a hexagonal cross section of size $d=34.4$ nm.}
 \label{Fig:InSbN75}
\end{center}
\end{figure}

\begin{figure}[t]
\begin{center}
\includegraphics[width=85mm]{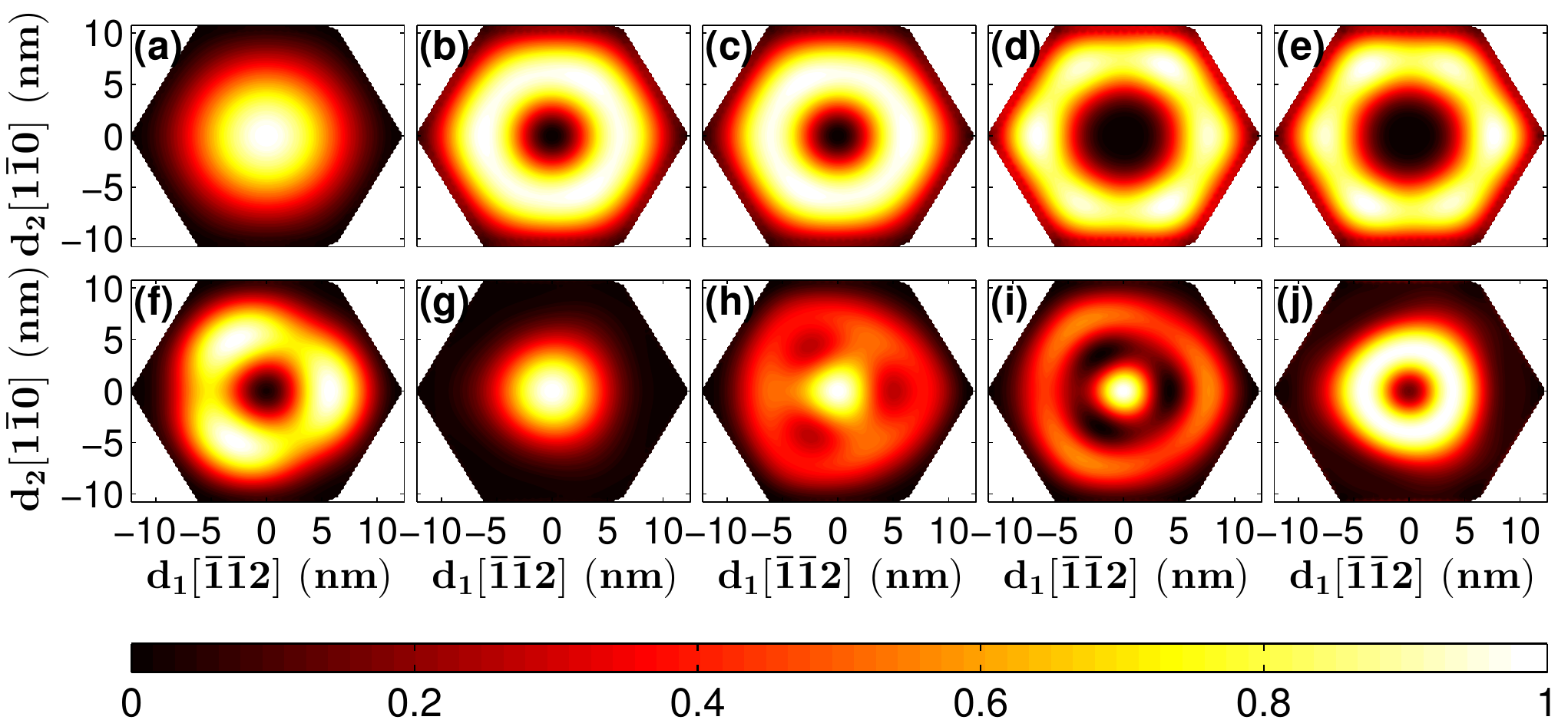}
\caption{(Color online) Wave functions of the five lowest conduction band states and the five highest valence band states at the $\Gamma$-point of the [111]-oriented GaSb nanowire with a hexagonal cross section of size $d=21.6$ nm. The wave functions are presented by the probability distributions on a (111)-plane of cation Ga atoms whose value at each cation atomic site is calculated by summing up the squared amplitudes of all the atomic orbital components on the cation atomic site and is normalized within each panel by the highest value found in the panel. Panels (a) to (e) show the wave functions of the five lowest conduction band states at the $\Gamma$-point, while panels (f) to (j) show the wave functions of the five highest valence band states at the $\Gamma$-point.}
 \label{Fig:GaSbN50}
\end{center}
\end{figure}

The wave functions of the [111]-oriented InSb and GaSb nanowires have also been calculated. The representative results are shown in Fig.~\ref{Fig:InSbN25} to Fig.~\ref{Fig:GaSbN50}. Here, a wave function is represented by the probability distribution on a (111)-plane of cation (In or Ga) atoms with the probability at each atomic site again calculated by summing up the squared amplitudes of all the atomic orbital components on that site and scaled against the maximum value within each graph. In these figures, the wave functions of the band states at the $\Gamma$-point are presented, at which all the states are double degenerate. Again, for each double degenerate state, we only plot the probability distribution for one of the two spin-degenerate states, since the other one has an identical spacial probability distribution. 

Figures~\ref{Fig:InSbN25} and \ref{Fig:InSbN75} show the calculated wave functions for the five lowest conduction band states and the five highest valence band states at the $\Gamma$-point of a [111]-oriented InSb nanowire with a hexagonal cross section of size $d=11.5$ nm and of a [111]-oriented InSb nanowire with a hexagonal cross section of size $d=34.4$ nm, respectively. It is seen that the corresponding conduction band states in the two InSb nanowires with very different sizes have the same spacial probability distribution characteristics. The lowest conduction band state (a $\Gamma_4$ state) in each InSb nanowire shows a highly symmetric, $s$-like wave function probability distribution and is very localized to the center region of the nanowire. This wave function characteristic may also be found in the calculations based on a simple one-band effective mass theory. The other four lowest conduction band states in each nanowire all show a donut-shaped probability distribution. Such a donut-shaped characteristic may not be obtained in the calculations for the wave functions of these states based on a simple one-band effective mass theory. Furthermore, these four conduction band states can be grouped into two groups based on their spacial localizations. The second lowest conduction band state (also a $\Gamma_4$ state) and the third lowest conduction band state (a $\Gamma_5$ or a $\Gamma_6$ state) are in one group and their wave functions are more localized around the center of the nanowire. The fourth lowest conduction band state (again a $\Gamma_5$ or a $\Gamma_6$ state) and the fifth lowest conduction band state (another $\Gamma_4$ state) comprise the second group and their wave functions are less localized to the center region of the nanowire. 

The wave functions of the corresponding valence band states of the two InSb nanowires with different sizes also show similar spacial distribution characteristics. Nevertheless, the distribution patterns of these valence band states look more complex than their counterpart conduction band states. Furthermore, it is interesting to see that the probability distribution of the highest valence band state (a $\Gamma_4$ state) of a [111]-oriented InSb nanowire is shaped as a donut with a $2\pi/3$-rotation symmetry, while the second highest valence band state (also a $\Gamma_4$ state) is a strong localized, $s$-like state inside the nanowire. The third highest valence band state (a $\Gamma_5$ or $\Gamma_6$ state) is also a localized state inside nanowire, but its wave function is shaped as two triangles of different sizes with a smaller solid one strongly localized at the cneter and a larger open one localized towards the surface of the nanowire. The fourth highest valence band state (again a $\Gamma_5$ or $\Gamma_6$ state) exhibits a donut-shaped probability distribution and is much more localized inside the nanowire when compared with the second to the fourth lowest conduction band states. The fifth highest valence band state (again a $\Gamma_4$ state) shows a similar probability distribution pattern as the third highest valence band state, i.e., it is shaped as two triangles of different sizes with a smaller solid one in the center region and a larger open one close to the surface of the nanowire. However, in comparison with the third highest valence band state, the wave function of the fifth highest valence band state exhibits a clear ring-like low probability structure in the region between the two high probability triangles. 

The wave functions of the conduction band states and the valence band states near the band gap of the [111]-oriented GaSb nanowires show similar probability distribution characteristics as the InSb nanowires. Therefore, in Fig.~\ref{Fig:GaSbN50}, we only present our calculated results for the wave functions of the five lowest  conduction band states and five highest valence band states at the $\Gamma$-point of the [111]-oriented GaSb nanowire with a hexagonal cross section of size $d=21.6$ nm. It is seen that the five lowest conduction band states at the $\Gamma$-point of the [111]-oriented GaSb nanowire have nearly the same probability distributions as the corresponding conduction band states of the two InSb nanowires shown in Figs.~\ref{Fig:InSbN25} and \ref{Fig:InSbN75}. The five valence band states at the $\Gamma$-point of the [111]-oriented GaSb nanowire also show the same probability distribution characteristics as their corresponding states in the InSb nanowires, except for some cases in which the ordering of the state probability distributions are interchanged. For example, the fourth and the fifth highest valence band states of the the [111]-oriented GaSb nanowire resemble the probability distributions of the fifth and the fourth higest valence band states of a corresponding InSb nanowire. However, when the symmetries of these states as shown in Figs.~\ref{Fig:[111]InSbband} and \ref{Fig:[111]GaSbband} are considered, the interchange in the ordering of the state probability distributions is seen to be fully consistent with the change in the symmetry ordering of these states.

\section{Conclusions}

In this paper, we present a theoretical study of the electronic structures of [001]-oriented free-standing InSb and GaSb nanowires with square and rectangular cross sections and of [111]-oriented free-standing InSb and GaSb nanowires with hexagonal cross sections. The band structures and the band state wave functions of these nanowires are calculated based on $sp^{3}s^{*}$, spin-orbit interaction included, tight-binding Hamiltonians  and the symmetry properties of the bands are analyzed in details based on double point groups. Generally, it is found that the conduction bands of these nanowires show good parabolic dispersion relations, while the valence bands show complex structures and orientation-dependent characteristics. Similarly, the wave functions of the conduction band states show simple, regularly ordered probability distribution patterns, while the wave functions of the valence band states show complex probability distribution patterns except for some of the top valence band states. In addition, although the wave functions of the band states in the [001]-oriented nanowires with a rectangular cross section of a large aspect ratio may be reproduced by a simple one-band effctive mass theory, such a simple theory could in general not be used to describe the probability distribution patterns of the conduction and valence band states of the [001]-oriented nanowires with a sqaure cross section and of the [111]-oriented nanowires with a hexagonal cross section. 

In details, it is shown that for the [001]-oriented InSb and GaSb nanowires with both a square and a rectangular cross section, all the energy bands are double degenerate, i.e., spin-degenerate. Thus, there is no spin splitting in the band structures of the [001]-oriented InSb and GaSb nanowires even in the presence of spin-orbit interaction in the band structure Hamiltonians. Nevertheless, for the [001]-oriented nanowires with a square cross section, the band structure Hamiltonians at the $\Gamma$-point are $D_{2d}$-symmetric, the energy band states at $\Gamma$-point are characterized by the $\Gamma_6$ and $\Gamma_7$ irreducible representations. In contrast,  for the [001]-oriented nanowires with a rectangular cross section, the band structure Hamiltonians at $\Gamma$ point are $C_{2v}$-symmetric, all the band states at the $\Gamma$-point are characterized by the $\Gamma_5$ irreducible representation. At a finite $k$ vector, the band structure Hamiltonians of the [001]-oriented nanowires with both a square and a rectangular cross section are $C_{2v}$-symmetric and thus all the band states are $\Gamma_5$-symmetric. It is also shown that for a [001]-oriented InSb and GaSb nanowires with a square cross section, even though the conduction bands show simple parabolic dispersion relations, the valence bands are characterized by anti-crossings and by a double maximum structure in the topmost valence band. In the case of the [001]-oriented nanowires with a rectangular cross section, the valence bands tend to show also good parabolic dispersions as the aspect ratio of the cross section is increased. The wave functions of the band states are characteristically different in the [001]-oriented nanowires with a square cross section and the [001]-oriented nanowires with a rectangular cross section. While the spacial probability distributions of the band states in the nanowires with a rectangular cross section show the characteristics that could well be described by a one-band effective mass theory, the spacial probability distributions of the band states in the nanowres with a square cross section show interesting structures with the characteristics that generally go beyond the predictions of a simple one-band effective mass theory.

For the [111]-oriented InSb and GaSb nanowires with a hexagonal cross section, the band structure Hamiltonians are $C_{3v}$ symmetric and the bands states are characterized by a two-dimensional $\Gamma_4$ representation and two one-dimensional $\Gamma_5$ and $\Gamma_6$ representations. Thus, not all band states are double degenerate (or spin degenerate) on the [111]-oriented nanowires, small but finite spin splittings occur in the band structures of these nanowires as a result of including spin-orbit interaction in the Hamiltonians. However, at the $\Gamma$-point, the full double degeneracy is restored and all the band ststes are double degenerate, i.e., spin degenerate, due to the presence of time reversal symmetry in the band structure Hamiltonians at this $k=0$ point. Again, the conduction bands of the [111]-oriented InSb an GaSb nanowires show good parabolic dispersion relations and the valence bands of these nanowires show complex dispersion relations. However, the valence bands of the [111]-oriented InSb and GaSb nanowires are dominantly characterized by band crossings, instead of anticrossings seen in the [001]-oriented nanowires with a square cross section, and the double maximum structure seen in the topmost valence band of a [001]-oriented nanowire with a square cross section does not fully develop in the [111]-oriented InSb and GaSb nanowires. Also, similarly as for the [001]-oriented nanowires with a square cross section, the wave functions of the band states in the [111]-oriented InSb and GaSb nanowires with a hexagonal cross section show the characteristic spacial distributions which go much beyond the prediction of a simple one-band effective mass theory. 

Finally, we have investigated the effects of quantum confinement on the band structures of the [001]- and [111]-oriented InSb and GaSb nanowires and have proposed an empirical formula for the description of quantization energies of the band edge states as a function of the lateral size in the nanowires. The formula can simply be used by experimentalists to estimate the enhancement of the band gaps of the nanowires as a result of quantum confinement. We believe that the results presented in this work will provide necessary information about the electronic structures of the [001]- and [111]-oriented InSb and GaSb nanowires and useful guidance for the practical use of these nanowires in the fabrications of novel nanoelectronic, optoelectronic and quantum devices. 

\section*{Acknowledgments}

This work was supported by the National Basic Research Program of China
(Grants No.~2012CB932703 and No.~2012CB932700) and the National Natural Science Foundation of China (Grants No.~91221202, No.~91421303, and No.~61321001).

\newpage

\end {document}